\documentclass[aps,pra,showpacs,twoside,twocolumn,longbibliography,10pt]{revtex4-1}
\usepackage[colorlinks=true, citecolor=red, urlcolor=blue ]{hyperref}
\usepackage{epsfig,newlfont,amssymb,amsfonts,amsmath,bm,subfigure,palatino,mathtools,amsthm,braket,times,soul,enumitem,color}
\usepackage[normalem]{ulem}
\newcommand{\stkout}[1]{\ifmmode\text{\sout{\ensuremath{#1}}}\else\sout{#1}\fi}
\usepackage[english]{babel}
\usepackage[utf8]{inputenc}
\usepackage{array}
\usepackage{xcolor}
\usepackage{graphics}

\def\Tr{\text{Tr}}
\newcommand{\kett}[1]{|#1\rangle}
\newcommand{\braa}[1]{\langle #1|}
\usepackage{amsthm}
\usepackage{verbatim}
\usepackage{bbm}
\usepackage{wrapfig}
\usepackage{makecell}

\usepackage{hyphenat}
    \usepackage{amsmath}

\newlength\figureheight 
\newlength\figurewidth 

\begin{document}

\title{Auxiliary-assisted energy distillation from quantum batteries}
\author{Paranjoy Chaki}
\affiliation{Harish-Chandra Research Institute,  A CI of Homi Bhabha National Institute, Chhatnag Road, Jhunsi, Allahabad 211 019, India}

\author{Aparajita Bhattacharyya}
\affiliation{Harish-Chandra Research Institute,  A CI of Homi Bhabha National Institute, Chhatnag Road, Jhunsi, Allahabad 211 019, India}

\author{Kornikar Sen}
\affiliation{Harish-Chandra Research Institute,  A CI of Homi Bhabha National Institute, Chhatnag Road, Jhunsi, Allahabad 211 019, India}

\author{Ujjwal Sen}
\affiliation{Harish-Chandra Research Institute,  A CI of Homi Bhabha National Institute, Chhatnag Road, Jhunsi, Allahabad 211 019, India}
\begin{abstract}
We discuss the idea of extracting energy from a quantum battery, applying a projective measurement on an auxiliary system. The battery is initially connected to the auxiliary system and allowed to interact with it. After some time, we execute a measurement on the auxiliary system
which probabilistically projects the setup to a particular state, and the corresponding state of the battery is the final state.
We consider the {sum of the} product of the energy difference between the initial and final states of the battery with the probability of getting that final state, where the sum is taken over all the preferable outcomes, that is, the outcomes which reduce the energy of the battery. We define the maximum value of this quantity as the {distillable energy}, where the maximization is taken over the time of interaction and auxiliary state and measurement basis parameters.
Restricting ourselves to a particular uncountable set of states, we find that {distillable energy} is always higher than the ergotropy of the battery, irrespective of the presence or absence of entanglement between battery and auxiliary.
%We show that a non-zero entanglement present initially between the battery and the auxiliary can induce an even higher amount of probabilistic energy extraction than that for product initial states. 
We also compare the {distillable energy} with the energy extracted using the interaction between the battery and the auxiliary, without any measurements.
{In comparison with the measurement-free scenario, we show that while measurement-based protocols do not provide any enhancement in the amount of extractable energy, they do yield a distinct advantage in terms of power, most notably in the case of distillable power, surpassing the power obtained without measurements.}
We show that the average of extracted energy overall measurement outcomes is independent of the applied measurement and equivalent to applying no measurement. Additionally, we find that the advantage in power for measurement-based energy extraction reduces with the increase in the size of the battery if the dimension of the auxiliary is kept fixed.  
Finally, we find that the ground state is the only state for which {distillable} energy is zero. {We also analyze the behavior of the maximum value (maximization done over the same parameters as in the case of distillable energy) of the product of the probability of getting a particular suitable outcome and the energy difference of the corresponding initial and final states of the battery, which is referred to as maximum probabilistically extractable energy.} 
\end{abstract}

\maketitle

\section{Introduction}
A battery is one of the most indispensable 
%electrical 
devices and has been widely used for centuries. It is essentially a system for storing energy that can be utilized whenever needed. These energy-preserving components are exploited in a variety of 
%electrical 
instruments in offices, homes, transports, and other places. Due to its widespread use, this device demands high portability and flexibility. Reducing its size to employ it in nanotechnology, such as nanochips, has been quite interesting in recent years. A decrease in the size of a battery greatly increases the potential of quantum mechanical features in it. For the past decade, research in the arena of such quantum batteries has led to profound results in quantum technology.

To our knowledge, R. Alicki and M. Fannes first formally introduced the idea of quantum batteries in an information-theoretic context by characterizing the maximum amount of energy that can be extracted from a quantum system, essentially the battery, by applying unitary operations~\cite{Alicki_2013}. The technical name assigned to this maximum extractable energy from a system is "ergotropy." Ergotropy is crucially related to the notion of passive states with respect to the unitary operation. States that do not have any energy that can be drawn from them by unitary operations are referred to as passive states. To get a more detailed understanding of passive states, we refer the reader to refs.~\cite{passive2,passive3,passive4,passive5,passive6,passive8,ref20}.

The charging power of a battery and quantum work capacitance~\cite{capacity,capacity2} are yet other elementary quantities that have the ability to judge how well a quantum battery performs. Many efforts have been made since R. Alicki and M. Fannes' work to determine the optimal charging and discharging protocols for quantum batteries and their charging power. Numerous models, including the short- and long-range XXZ quantum spin chains~\cite{xyz2}, spin cavities~\cite{cavitya,cavityb,cavityc,cavityd,cavitye,cavityf,cavityg,decay,cavityh1}, non-Hermitian systems~\cite{non-hermitian}, bose- and Fermi-Hubbard, as well as their disordered versions~\cite{PhysRevA.106.022618}, have been explored to determine the best quantum battery with respect to their performance. A dimensional enhancement in the charging of quantum batteries has been witnessed~\cite{Ghosh_2022}. Adding disorder is also proven to facilitate the charging process of open quantum batteries~\cite{Ghosh_2020}. The behavior of extractable energy in the localization phase of many-body quantum chains has been discussed in Refs.~\cite{localised1,localization2}.
There exist various research works that explore how quantum resources~\cite{resource1}, for example, entanglement~\cite{Entt_1,ent2,demon1} and quantum coherence~\cite{coh_1,coh_2,coh_3,coh_4}, can influence the charging and discharging strengths of quantum batteries. Charging and discharging of open or noisy quantum batteries are also interesting and popular fields for research. Important work has been done in this direction, such as in~\cite{opena,open2,openb,openc,opend,Ghosh_2021,opene,openg,openh,IIII}. An extensive comparison between quantum and classical many-body batteries can be found in Ref.~\cite{classical}.
Quantum batteries have been successfully implemented in experiments~\cite{exp4,exp3,exp2,exp1}.
%Quantum batteries have also been experimentally implemented in~\cite{exp1,exp2,exp3,exp4}.

In this paper, we mainly focus on the process of energy extraction. A unitary operation is the most traditional tool used for energy extraction. A quantum battery is typically described by a state, which specifies the system, and a Hamiltonian, which defines its energy. In the conventional method of energy extraction, a certain field is added to the system Hamiltonian for some fixed amount of time, which causes the system to evolve unitarily. As a result of the combined effects of the applied field and the pre-existing system Hamiltonian, the energy of the battery is extracted.

In Ref.~\cite{Yan_2023}, the authors have introduced a new technique for charging quantum batteries, viz. by combining an auxiliary system with the battery and then performing a sequence of measurements on the auxiliary system. A finite time gap has been considered between each consecutive measurement, within which the battery and the auxiliary systems are left to evolve. In this paper, we use this idea of performing measurements and utilize it in the context of the extraction of energy. The basic idea is that we attach an auxiliary system to the battery. Then the composite system, which includes the battery and the auxiliary systems, is allowed to evolve under a joint Hamiltonian. The Hamiltonian is composed of the system's and the auxiliary's local Hamiltonians and a global Hamiltonian characterizing the interaction between them. In order to extract energy, a projective measurement is performed on the auxiliary system at a desired moment. A specific outcome of the measurement is chosen, which may occur with a non-zero probability. The auxiliary is then ignored, and the corresponding state of the battery is considered to be the final state. 
{
The relevant quantity of interest is the product of the probability associated with the preferred measurement outcome and the corresponding energy difference between the initial and final states of the battery corresponding to that outcome. Since our focus is on energy extraction via the measurement-based protocol, we prefer only those outcomes for which the energy difference is positive. Finally, we take a sum of this quantity over all such favorable outcomes and maximize the sum over the initial state of the auxiliary, the time of evolution, and the applied measurement. The resultant quantity is dependent only on the state of the battery and is referred to as the "distillable energy" of the battery. This protocol is inspired by other distillable protocols, such as entanglement distillation~\cite{bbpssw,bennett_distil}, coherence distillation~\cite{coh_distil1,coh_distil2,coh_distil3}, magic distillation~\cite{magic_distil1,magic_distil2,magic_distil3,magic_distil4,magic_distil5} etc. For example, in entanglement distillation, non-maximally entangled states are probabilistically converted into a smaller number of maximally entangled states through a process involving post-selection and a success probability less than unity. 
In entanglement distillation, specific measurement outcomes that satisfy the success criteria of the protocol are retained, while all other outcomes are discarded.
%in  entanglement distillation, particular measurement outcomes are retained which are fruitful according to the criteria of the protocol and  other outcomes are discarded. 
This notion of outcome-dependent post-selection similarly underlies our measurement-based energy extraction scheme, thereby motivating the use of the term distillable energy.
}
%Similar probabilistic distillation protocols have also been extensively studied in the context of quantum coherence~\cite{coh_distil1,coh_distil2,coh_distil3} and magic state distillation~\cite{magic_distil1,magic_distil2,magic_distil3,magic_distil4,magic_distil5}.}
%Similar probabilistic protocols are also well-known in coherence~\cite{coh_distil1,coh_distil2,coh_distil3} or magic distillation~\cite{magic_distil1,magic_distil2,magic_distil3,magic_distil4,magic_distil5}.}
%%%%%%%%%%%%%%%%%%%%%%%%%%%%%%%%%
%\textcolor{blue}{The product of the probability of getting that outcome and the energy difference between the initial and final states of the battery can be defined as the probabilistically extracted energy through measurement.} 
%%%%%%%%%%%%%%%%%%%%%%%%%%%%%%%%%%%
{ 
Alternatively, one may consider a different figure of merit defined as the product of the probability of obtaining a specific measurement outcome and the corresponding energy difference between the initial and final states of the battery, that is, instead of considering the sum of all relevant outcomes, only a single outcome can be chosen. Upon maximizing this quantity over the initial state of the auxiliary system, the time of measurement, and the choice of measurement basis, we refer to the resulting optimized value as the maximum probabilistically extractable energy via measurement. Clearly, the maximum probabilistically extractable energy will always be less or equal to the distillable energy.}
%%%%%%%%%%%%%%%%%%%%%%%%%%%%%%%%%
%\textcolor{blue}{Accepting the non-deterministic nature of measurement outcomes, we refer to this optimal probabilistic energy difference as probabilistically extractable energy.} 
%%%%%%%%%%%%%%%%%%%%%%%%%%%%%%%%%
%\textcolor{magenta}{Magic distillation}~\cite{magic_distil1,magic_distil2,magic_distil3,magic_distil4,magic_distil5}, \textcolor{magenta}{entanglement distillation, \textcolor{magenta}{coherence distillation}~\cite{coh_distil1,coh_distil2,coh_distil3}

{Our first motive is to compare this technique with the conventional energy extraction method, which uses only unitary operations on the battery. Not only do we determine that both of these measurement-based protocols provide more energy than unitary operations, but we also witness that there exist passive states with respect to the unitary operation from which, though energy extraction by unitary operations is not possible, the measurement-based protocols can extract a significant amount of energy. Moreover, we show that the performance time of the measurement-based energy extraction is much less than the unitary-based method. The optimal parameters which provide the maximum extractable energy are analyzed. 
%\textcolor{blue}{We also make a comparison between the distillable energy with that of the process where only the optimal time-dependent global unitary acts on the battery and auxiliary to extract energy, but no measurement is being performed. }
%\textcolor{blue}{
%\textcolor{violet}{We find that, in the absence of any measurement on the auxiliary system, the presence of initial entanglement between the battery and the auxiliary system offers no advantage compared to the case without initial entanglement.}
%We have found that no advantage is present when initial entanglement is present between the battery and auxiliary system over the situation where there is no initial entanglement, in the scenario where no measurement is performed on the auxiliary system. }
%We have \textcolor{green}{seen} that no advantage is present when initial entanglement is present between the battery and auxiliary system where no measurement is performed on the auxiliary system. 
%The situation is true for distillable energy as well. However the initial entanglement between battery and auxiliary system is 
%beneficial for maximum probabilistically extractable energy.} 
Further, we show that if a probabilistic average of the extracted energy is performed over all measurement outcomes, then this average extracted energy becomes independent of the applied measurement and becomes equal to the difference between the energies of the initial and final evolved state of the battery, just before the application of the measurement.
 Moreover, in comparison with the measurement-less scenario, where only the global unitary acts on the battery and auxiliary system, we show that although measurement-based protocols do not exhibit any advantage in terms of extractable energy, they do offer a clear advantage in terms of power, particularly for the distillable power over the powers where no measurement is involved. In this regard, we have also seen that no advantage is present when initial entanglement is present between the battery and auxiliary system in a measurement-less scenario. The situation is true for distillable energy as well. However, the initial entanglement between the battery and the auxiliary system is
beneficial for maximum probabilistically extractable energy.
 Furthermore, we find that increasing the size of the battery decreases the advantage of the measurement-based energy extraction method when the auxiliary's size is kept fixed. We also show that, although measurement-based energy does not offer an advantage over the measurement-free scenario for the considered battery sizes, the power associated with the measurement-based protocol provides benefits over the measurement-less scenarios, across all battery sizes. Finally, we try to collect the states from which even the measurement-based technique is unable to extract any energy; that is, we  find the set of measurement-passive states (MPS). We determine that the set only consists of a single element, the ground state of the considered Hamiltonian.}

The main difference between Ref.~\cite{Yan_2023} and our work, apart from the fact that here we focus on energy extraction instead of charging, is that, our protocol includes only one measurement on the auxiliary qubit and not a sequence of measurements. Moreover, our motive in this work is to examine the fundamental notions involved in energy extraction, namely the dependence of the extracted energy on the initial entanglement content between the battery and the auxiliary, the set of MPS states, etc. There are also some other studies on quantum batteries where measurements are performed to gain an advantage~\cite{demon1,coh_1,coh_2,homo_1,demon2}. But the energy extraction protocols discussed in those works are undoubtedly different from the ones presented in Ref~\cite{Yan_2023} as well as in this paper. Recent NMR experiments focus on estimating energy and ergotropy during discharging  quantum battery~\cite{mahesh}. Various works using weak~\cite{nmr_w} and projective measurements~\cite{proj1,proj2} with post-selected outcomes have made the discharging of an NMR quantum battery a stochastic process, showcasing probabilistic results in quantum technology.

{The remainder of the paper is structured as follows: We present a detailed description of our protocol in the first part of Sec. \ref{22}. The second and last part of Sec. \ref{22} contains two subsections. In the first subsection, \ref{a}, we consider the case where the initial joint states of the battery and the auxiliary are separable. A discussion on the parameters which optimize the protocol, and on the speed of the process is provided in the same subsection. The other subsection, viz. \ref{b}, focuses on the situation where there exists a finite non-zero amount of entanglement between the battery and the auxiliary system initially. We have also shown the oscillatory behavior of distilled energy and probabilistically extractable energy for both initially product and entangled states. In Sec.~\ref{3x1}, we make a comparison between distillable energy and daemonic ergotropy.
In Sec.~\ref{s4cx}, we determine the extractable energy from the battery, without performing any measurement, just by switching on the interaction between the battery and the auxiliary, which is used in the measurement-based method. After that we also observe how the maximum power behaves in both measurement-based and without-measurement scenarios. Sec.~\ref{newsec2} shows measurement does not affect the average extracted energy, averaged over all possible measurement outcomes. The performance of the measurement-based energy extraction method with multi-qubit batteries is analyzed in \ref{newsec3}. In Sec.~\ref{3}, we depict the set of states corresponding to the considered Hamiltonian from which no energy can be extracted using the measurement-based protocol. We provide the concluding remarks in Sec.~\ref{concl}.}

\section{Energy extraction protocols and comparisons between them}
\label{22}
A quantum battery is described by a state, $\rho_B$, and a Hamiltonian, $H_B$. In recent years, a large amount of research has focused on finding the optimal methods for charging and discharging quantum batteries. Let us first describe the most common method of energy extraction from them.\\
\textbf{Unitary-based protocol}~\cite{Alicki_2013}\textbf{:} The conventional approach to energy extraction is to unitarily evolve the system. A certain potential can be introduced depending on whether one wants the system to be charged or discharged.

Let us assume that $V(t)$ is a time-dependent potential that can be employed for energy extraction and that it has been switched on for the duration of time $T$. The initial state of the battery, just before adding the field, can be denoted by $\rho_{B}$. After time $T$, the battery will be in the state $U(T)\rho_{B}U^\dag(T)$, where
\begin{equation}
    U(T)=\mathcal{T}\exp{\left(-i \int_{0}^{T} dx [H_B+V(x)]/\hbar\right) }.\nonumber\end{equation}
Here $\mathcal{T}$ denotes the time ordering. The amount of energy extractable through the unitary, $U(T)$, from the state, $\rho_B$, is given by
\begin{equation}
    W=\Tr(\rho_{B} H_B)-\Tr\left[U(T)\rho_{B}U^{\dag}(T) H_B\right].\label{2}
\end{equation}
%$W$ is defined as the amount of energy that the field $V(T)$ extracts during the course of the time interval, $T$.
By selecting the optimal unitary, which minimizes the second term of Eq. \eqref{2}, the energy extraction can be optimized. The maximum amount of extractable energy can thus be written as 
\begin{eqnarray}
    W^U_{}&=&\Tr(\rho_{B} H_B)-\min_{U}\Tr(U\rho_{B}U^{\dag} H_B)\nonumber\\
    &=&\Tr(\rho_{B} H_B)-\Tr(U_{\min}\rho_{B}U_{\min}^{\dag} H_B),\label{3}
\end{eqnarray}
where $U_{\min}$ is the unitary operation for which the optimum value can be achieved. {Since, in the next section, we will discuss energy extraction using measurements, to specifically denote the extractable energy using only unitary operations, we use the symbol
$W^{U}$, where
$U$ in the superscript of $W$ denotes unitary operations in the left-hand side of the above expression.}
%From the state $\rho_\sigma=U_{\min}\rho_{int}U^{\dag}_{\min}$ no energy can be extracted. These states are known as passive states. The formal definition of passive state, $\sigma$, is that they are the states from which no energy can be extracted using unitary operations. Passive states always commutes with the system Hamiltonian, say $H$. Even more the eigenvalues of $\sigma$ are non-increasing with the eigenvalues of $H$. That means for a particular state $\rho_{int}$ and Hamiltonian $H$ if the eigenvalues passive state $\sigma_{\rho,H}$  $\lambda_1<\lambda_2<\lambda_3<.......<\lambda_n$then eigenvalues of the Hamiltonian $H$ will be in a order  $ e_1>e_2>e_3>.......>e_n$. The energy to this passive state is given by $\lambda_ie_i$. Now, the energy extraction with respect to the state $\sigma_{\rho,H}$ is,

A state, $\sigma$, corresponding to a Hamiltonian, ${H}$, is called passive with respect to the unitary operation, if no unitary operation can extract energy from the state, $\sigma$. Thus for the states, $\sigma$, we can write
$$\text{Tr}(\sigma H)\leq\Tr(U\sigma U^\dagger H)\text{, for all unitary operations, }U.$$ Let the set of eigenvalues and eigenvectors of $H$ be $\{\epsilon_i\}_i$ and $\{\kett{\epsilon_i}\}_i$, respectively,  where the eigenvalues are arranged in ascending order, i.e., $\epsilon_i\leq\epsilon_{i+1}$ for all $i$. Then passive state, $\sigma$, corresponding to the Hamiltonian, $H$, is known to commute with $H$ and can always be expressed in the form $\sigma=\sum_i \lambda_i \kett{\epsilon_i}\braa{\epsilon_i}$ where $\lambda_i\geq \lambda_{i+1}$ are the set of eigenvalues arranged in decreasing order. Precisely, if for a particular arrangement of the basis, $\{\kett{\epsilon_i}\}$, the eigenvalues of the Hamiltonian satisfy $\epsilon_i<\epsilon_j$, then the eigenvalues of the passive state with respect to the unitary operation will follow the opposite order, i.e., $\lambda_i\geq \lambda_j$~\cite{Lenard_1978,Pusz_1978}.

Corresponding to every initial battery state, $\rho_{B}$, there exists a passive state, $\sigma_{\rho_{B}}$, and a unitary operator, $U_{\min}$, such that $\sigma_{\rho_{B}}=U_{\min}\rho_{B} U_{\min}^\dagger$. The unitary's suffix signifies that it is the unitary that can extract the most possible energy from the battery, initially prepared in the state $\rho_{B}$. Because once one reaches the passive state, $\sigma_{\rho_{B}}$, no further energy can be extracted from it. Thus, the maximum amount of energy that can be extracted from a state, $\rho_{B}$, is 
\begin{equation}
W_{}^U(\rho_{B},H_B)=\text{Tr}(\rho_{B} H_B)-\text{Tr}(\sigma_{\rho_{B}} H_B). \label{teq2}
\end{equation}

\textbf{Measurement-based protocol:}\label{mp} In this paper, we discuss a energy extraction technique that can outperform the previously mentioned unitary-based protocol. For example, we show that this protocol can squeeze out energy even from most passive states with respect to the unitary operation.% Therefore, a maximally mixed state do not act as a passive state. During a typical energy extraction process, a maximally mixed state acts passively. Here, we provide a method of energy extraction that permits the utilisation of even highly mixed states for energy extraction. 

This energy extraction method involves measurement. In particular, our aim here is to investigate how the measurement performed on an attached entity can be used to extract energy from the battery. In this regard, we consider the simplest situation involving only a single measurement on an auxiliary qubit.

To formulate the protocol, let us consider a single-qubit battery, $B$, and a single-qubit auxiliary system, $A$. The initial joint state of the battery and auxiliary qubits can be denoted as $\rho_{BA}(0)$, which acts on the composite Hilbert space $\mathcal{H}_B\otimes\mathcal{H}_A$. We fix the Hamiltonians describing the energies of $B$ and $A$ to $H_B$ and $H_A$, respectively, and defined as 
\begin{equation}
H_B=h\sigma^z\text{ and }H_A=h\sigma^z.\label{4eq1}
\end{equation}
 Let us assume that the auxiliary and battery are brought together and an interaction, $H_I=J(\sigma_x \otimes \sigma_x)$, is switched on. Then the total Hamiltonian of the composite system consisting of $B$ and $A$ is given by
%Let us consider a single-qubit battery, $B$. The initial state of the battery, at time $t=0$, can be represented by the density operator $\rho_B$ which acts on the Hilbart space $\mathcal{H}_B$. We additionally take into account another single-qubit auxiliary system, $A$, that is detached from the battery, and prepared in a separate state, $\rho_A$, which acts on the Hilbert space $\mathcal{H}_A$.  
%Thus the joint state of the battery and auxiliary qubits is given by $\rho_{BA}(0)=\rho_B\otimes\rho_A$. We fix the Hamiltonians describing the self energies of $B$ and $A$ as $H_B=h\sigma^z$ and $H_A=h\sigma^z$, respectively. Let us assume that the auxiliary and battery can be braught together and an interaction, $H_I=J(\sigma_x \otimes \sigma_x)$, can be switched on. Then the total Hamiltonian of the composite system consisting $B$ and $A$ is given by,
\begin{eqnarray}\label{hamiltonian}
H_{BA}&=&H_B+H_A+H_I\nonumber\\&=&h(\sigma^z\otimes I_2)+h(I_2\otimes \sigma^z)+J(\sigma^x \otimes \sigma^x). \label{4eq5}                  \end{eqnarray}
We use the notations $\sigma^x$, $\sigma^y$, and $\sigma^z$ to describe the three Pauli spin operators and $I_2$ to represent the identity operator acting on the qubit Hilbert spaces. For all the numerical calculations, we will consider $J=2h$.

The state of the entire system after time $t$ can be expressed as 
    \begin{eqnarray}
        \rho_{BA}(t)&=&U_H(t)\rho_{BA}(0)U_H(t)^\dag,\nonumber\\ \text{where } U_H(t)&=&\exp(-iH_{BA}t/\hbar).
        \label{4eq3}
         \end{eqnarray}
         
         	\begin{figure}
		\centering
	\includegraphics[scale=0.12]{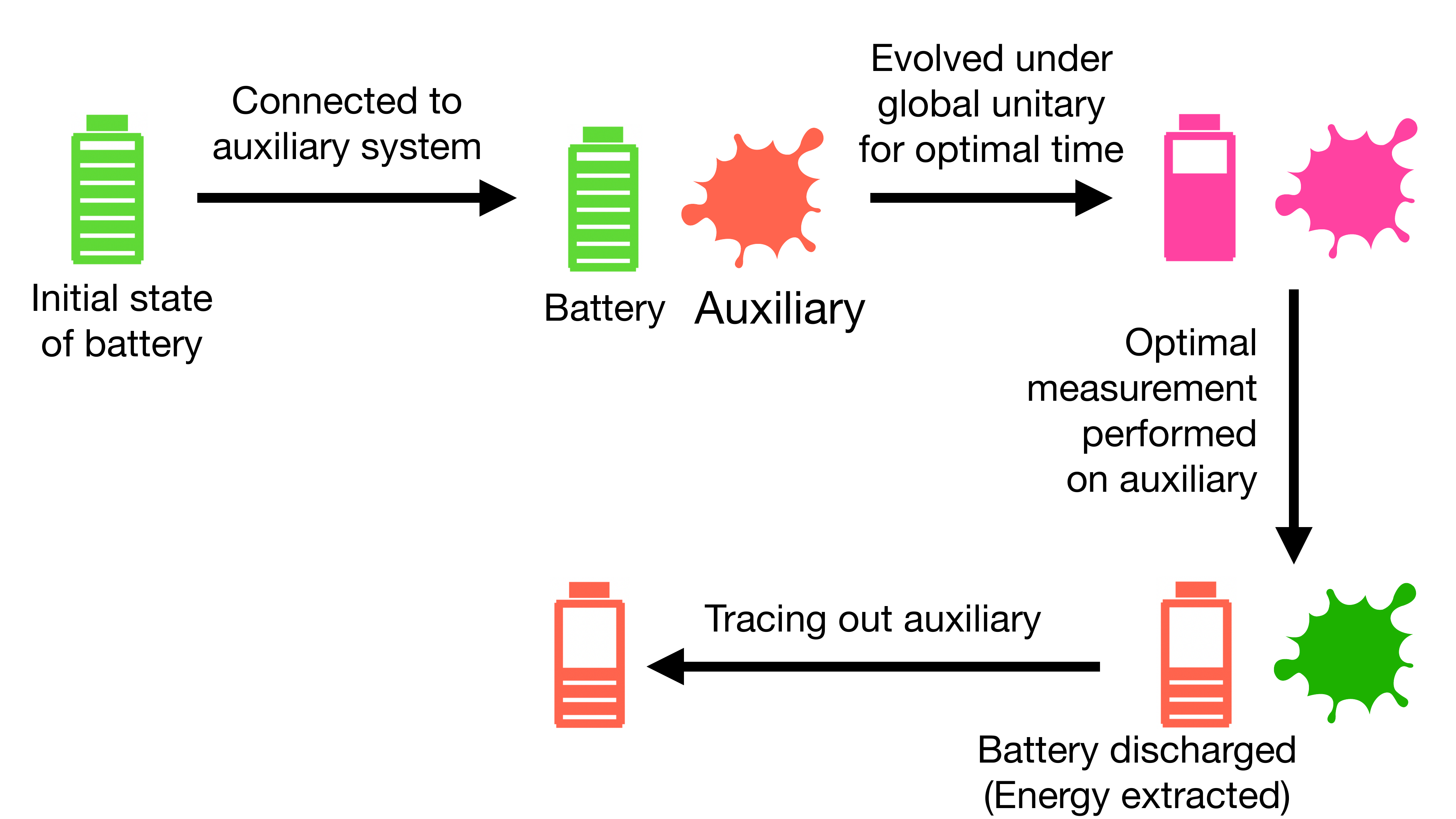}
		\caption{Schematic representation of auxiliary-assisted measurement-based energy extraction procedure.
   The green battery in the diagram represents a single-qubit quantum battery. After connecting the battery to an auxiliary, the entire battery-auxiliary system is evolved under a unitary evolution. An optimal projective measurement is then carried out on the auxiliary system for which the extractable energy would be maximum. The auxiliary is then traced out. It should be mentioned that in order to increase the amount of extractable energy, we additionally optimize the process over the time instance at which the measurement is performed and the initial state of the auxiliary. }
		\label{fig:my_label}
	\end{figure}
To extract energy from the battery, $B$, after a certain time $t$, we perform a rank-1 projective measurement on the auxiliary qubit, $A$, in an arbitrary but fixed basis $\{\kett{\psi},\kett{\psi_{\perp}}\}$. The normalized orthogonal basis can be expressed using two real parameters, $\theta$ and $\phi$, in the following way:
    \begin{eqnarray}
        \kett{\psi}&=&\cos(\frac{\theta}{2})\kett{0}+\exp(-i\phi)\sin(\frac{\theta}{2})\kett{1}\nonumber\nonumber\\
        \kett{\psi_{\perp}}&=&\sin(\frac{\theta}{2})\kett{0}-\exp(-i\phi)\cos(\frac{\theta}{2})\kett{1}\label{neweq1},
    \end{eqnarray}
where $\theta\in[0,\pi]$ and $\phi\in[0,2\pi)$. Here, $\kett{0}$ and $\kett{1}$ are the excited and ground states of $\sigma^z$, respectively. {After performing the measurement, we select one of the outcomes and ignore (mathematically, trace out) the auxiliary component.
Let the battery state corresponding to the chosen outcome, $i$, after tracing out the auxiliary, be $\rho^i_B$. Hence, the energy difference between the initial and final state of $B$ is
\begin{equation*}
    \Delta E^i=\Tr(H\rho_B)-\Tr(H\rho^i_B),
\end{equation*}
where $\rho_B=\Tr_A\left[\rho_{BA}(0)\right]$. Clearly $\Delta E^i$ depends on the basis on which the measurement has been done as well as the chosen outcome.}

{We know in quantum mechanical systems, the result of an applied measurement is a probabilistic event, i.e., each outcome occurs with a certain probability. Let us assume the probability of getting the $i^{\text{th}}$ outcome is $P_{o}^i$. The $i^{\text{th}}$ measurement operator transforms the state of $B$ to $\rho_B^i$. Hence, even if the amount of extracted energy, $\Delta E^i$, from a battery corresponding to a particular outcome is very high, the probability, $P_o^i$, of getting that outcome can be very small. To take into account the non-deterministic nature of measurements, we multiply the extracted energy ($\Delta E^i$) with the probability ($P_o^i$) of the corresponding outcome and focus on the whole quantity, $P_o^i\Delta E^i$.
We take the sum of all such terms in which the outcomes correspond to a decrease in the energy of the battery.}
%If $P_o\Delta E$ is very high we know both extractable energy, $\Delta E$, and also the probability, $P_o$, of getting that energy is significantly large. We refer to this quantity as probabilistically extracted energy and denote it as $W_{P}$.}
%We know in quantum mechanical systems, the result of an applied measurement is a probabilistic event, i.e., each outcome occurs with a certain probability. Let us assume the probability of getting the particular outcome, which would transform the state of $B$ to $\rho_B'$, be $P_{o}$. Hence, even if the amount of extracted energy, $\Delta E$, from a battery corresponding to a particular outcome, is very high, the corresponding probability, $P_o$, of getting that outcome can be very small. To take into account the non-deterministic nature of measurements, we multiply the extracted energy ($\Delta E$) with the probability ($P_o$) of the corresponding outcome and consider the whole quantity ($P_o\Delta E$) as our figure of merit. If $P_o\Delta E$ is very high we know both extractable energy, $\Delta E$, and also the probability, $P_o$, of getting that energy is significantly large. We refer to this quantity as probabilistically extracted energy and denote it as $W_{P}$.
{Finally, we maximize this sum over the total time of the unitary evolution, $t$, the measurement basis, $\{\kett{\psi},\kett{\psi^\perp}\}$, and the joint initial state $\rho_{BA}(0)$, keeping the initial state of $B$, $\rho_B$, and the total Hamiltonian of the composite system, $\bar{H}_{BA}$, fixed, and name it as distillable energy. The mathematical definition of distillable energy is as follows:
        \begin{equation}
W^{D}(\rho_B,H_B)=\max_{t, \theta, \phi, \rho_{BA}(0)}\sum_{i\in \mathbb{I}}P_{o}^i\Tr\left[\left(\rho_B-\rho_B^i\right)h\sigma_z\right], 
    \end{equation}
    where the summation runs over the set of all outcomes, $\mathbb{I}$, for which the change in energy of the battery is positive, i.e., $\Tr\left[\left(\rho_B-\rho_B^i\right)h\sigma_z\right]> 0$.
    In Fig.~\ref{fig:my_label}, we present a schematic diagram of the protocol.} 

{Instead of summing over all $i\in\mathbb{I}$, one can also consider the quantity $P_{o}^i\Tr\left[\left(\rho_B-\rho_B^i\right)h\sigma_z\right]$, corresponding to a single outcome, $i$, for a given measurement setting, given auxiliary system, and given time $t$, and termed as probabilistically extractable energy. Now if we maximize the probabilistically extractable energy over all possible measurement settings, auxiliary systems, and time $t$, then the maximum probabilistically extractable energy is mathematically given by
\begin{eqnarray}
    W^{M}(\rho_B,H_B)=\max_{t, \theta, \phi, \rho_{BA}(0)}P^i_{o}\Tr\left[\left(\rho_B-{\rho_B^{i}}\right)h\sigma_z\right], 
\end{eqnarray}
where the $M$ in the superscript helps to keep in mind that the energy extraction process here is measurement-based.}
    %In Sec.~\ref{sec_new}, we discuss why such an averaging is irrelevant in the context of energy extraction from batteries by performing projective measurement on an auxiliary attached to the battery.}   

{We would like to mention here that in order to utilize a quantum battery for energy extraction, we need to know how much energy is stored in the battery initially. For that, we need to measure the initial energy of the battery. Therefore, for an unknown battery, an initial measurement is required. Additionally, to extract energy, another measurement is being implemented in our protocol. Therefore, a two-time measurement scheme \cite{two-point} is involved here, one for the characterization of the stored energy and the other for energy extraction. Here, however, we assume that this characterization has already been done by measuring on the ideal initial copy of the battery state. So we move on with the known initial quantum state, $\rho_B$, and focus on the measurement performed on the auxiliary state to extract energy.} 

% We also want to mention here that since measurement outcomes are probabilistic and our measurement-based energy extraction involves post-selection, the amount of energy that can be extracted in this method is probabilistic. Instead of averaging overall outcomes here, we have chosen a particular optimum outcome which will appear with a probability. Instead of considering a particular outcome, one may also consider the average energy extraction, averaged over all measurement outcomes. 
We also want to mention here that since measurement outcomes are probabilistic and our measurement-based energy extraction involves post-selection, the amount of energy that can be extracted in this method is probabilistic. Instead of averaging over all outcomes, here we chose all such optimum outcomes for which the energy of the battery reduces. We sum over all such preferred outcomes. Instead of considering this protocol, one may also consider the average energy extraction, averaged over all measurement outcomes. In Sec. \ref{newsec2}, we show the average extractable energy, with the average being performed over all measurement outcomes of the projective measurement performed on the auxiliary qubit, is independent of the direction of measurement and is always equal to the energy difference between the states of $B$ before and after its interaction with the auxiliary.

{We want to find out if there are any instances in which $W^D(\rho_B,H_B)$ and  $W^M(\rho_B,H_B)$ are larger than $W^U(\rho_B,H_B)$. In this regard, in subsection \ref{a}, we will consider the initial joint state of the system and the auxiliary as a product. After realizing that $W^D(\rho_B,H_B)$ and $W^M(\rho_B,H_B)$ are always greater than $W^U(\rho_B,H_B)$ for our considered model, we will move on to examine if the extractable energy in measurement based protocols can be enhanced further by considering shared entanglement in the initial state of the auxiliary and battery in Subsection \ref{b}. We also include a table (Table~\ref{t1}) where we mention all the figures of merit, their notations, and their corresponding colors and types in the plots.}

\begin{table*}
    \centering
    \begin{tabular}{|c|c|c|c|c|}
        \hline
        \textbf{Types of Figure of Merits} 
        & \multicolumn{2}{c|}{\textbf{Symbols }} 
        & \multicolumn{2}{c|}{\textbf{Colours of curves}} \\
        \hline
        & \textbf{Initially product} & \textbf{Initially entangled} & \textbf{Initially product} & \textbf{Initially entangled}  \\
        \hline
        Distillable energy 
        & $W^{D}_S$ & $W^{D}_E$ & Pink (dot) & Dark slate green (smooth curve) \\
        \hline
        Maximum probabilistically   
        & $W^{M}_S$ & $W^{M}_E$ & Green (smooth curve) & Red (smooth curve)\\extractable & & & &
         \\energy & & & & \\
        \hline
                Ergotropy 
        & $W^{U}$ & $W^{U}$ & Blue (smooth curve) & Blue (smooth curve) \\
        \hline
        Maximum extractable energy  
        &  $W^{U}_S$ & $W^{U}_E$ & Blue (dot) & Yellow (stars) \\
         without measurement & & & &\\
         \hline
        Distillable power 
        & $P^{D}_S$ & $P^{D}_E$ & Brown (dot) & Gray (dot) \\
        \hline
        Maximum probabilistic power 
        & $P^{M}_S$ & $P^{M}_E$ & Crimson (dot) & Purple (dot) \\
            \hline
        Power without a measurement 
        & $P^{U_H}_S$ & $P^{U_H}_E$ & Teal (square box) & Yellow (square box) \\
        \hline
        Distillable energy from & $W_N^D$ &-& \makecell{Orange (dot) with navy blue border}&-\\ $N$ qubit battery  & & & &\\
        \hline
        Maximum probabilistically  & $W_N^M$ &-& \makecell{Blue (dot)\\ with navy blue border}&-\\  extractable energy & & & & \\ from $N$ qubit battery  & & & &\\ 
        \hline
        Maximum extractable  & $W_N^{U_H}$ &-& Yellow (star)&-\\ measurement from $N$  & & & &\\energy without & & & & \\ qubit battery  & & & &\\
        \hline
         Distillable power from & $P_N^D$ &-& Light coral (hexagon)&-\\ $N$ qubit battery  & & & &\\
         \hline
           Maximum  & $P_N^M$ &-& Wheat (hexagon)&-\\ probabilistic power & & & & \\ from $N$ qubit battery  & & & &\\ 
        % \hline
        %        Maximum power without & $P_N^{U_H}$ &-& Dark slate blue with navy blue border (star)&-\\ measurement from $N$  & & & &\\ qubit battery  & & & &\\
         \hline
         Maximum power without & $P_N^{U_H}$ & - & \makecell{Dark slate blue with \\ navy blue border (star)} & -\\
      measurement from $N$  & & & &\\ 
         qubit battery  & & & &\\
           \hline
         average extractable energy 
        & $W_{av}^M$ & $W_{av}^M$ & - & - \\
        \hline
    \end{tabular}
    \caption{{Here we  provide the list of the relevant figures of merit used throughout the paper along with their notational symbols and colors used in the figures to denote the respective quantities. For instance, ergotropy for a given marginal state is independent of whether it is entangled or not. Therefore, for both the cases, we use the same color, i.e. blue.}}
    \label{t1}
\end{table*}

%and, we presume that the battery's initial state is entangled with the auxiliary. Finally we will compare the ergotropy of the battery with the amount of probabilistic energy extraction considering both of these scenarios. 

\subsection{Case I: Initial battery-auxiliary state is separable}
\label{a}
%We are interested in focusing on the initial composite state of the battery and auxiliary, which, in this subsection, will be considered to be product. 
Here, we compare the ergotropy of a battery with the extractable energy from the battery using the introduced measurement-based energy extraction method. In this regard, we consider the composite state of $A$ and $B$ to be a product.
Let us denote the joint initial state of $BA$ as $\rho_{BA}=\rho_B \otimes \rho_A$. We consider the initial battery state to be
\begin{equation}
    \rho_B=\left(\frac{1+k}{2}\right)\ket{0}\bra{0}+\left(\frac{1-k}{2}\right)\ket{1}\bra{1}\nonumber,\text{ where }-1\leq k\leq 1.
\end{equation}
Expressing the battery's state in the $\{\kett{0},\kett{1}\}$ basis, we have
\begin{equation}
    \rho_B=\begin{bmatrix}
			\frac{1+k}{2} & 0 \\
			0 &\frac{1-k}{2} \\
		\end{bmatrix}\label{4eq2}.
    \end{equation}
   The battery's state is assumed to be devoid of any quantum coherence~\cite{aberg2006quantifying, PhysRevLett.113.140401, PhysRevLett.116.120404, RevModPhys.89.041003}, in the local energy eigenbasis, as we wish to understand the effect of entanglement of the battery with the auxiliary in the 
   %probabilistically extractable
   {measurement-based energy extraction}, in the absence of the other paradigmatic quantum resource, viz., quantum coherence. The ergotropy of such a battery is
    \begin{eqnarray}
        W^U(\rho_{B},H_B)&=&2hk \text{ for }k\geq 0\text{ and }\\
        &=&0\text{ otherwise.}
        \label{4eq4}
    \end{eqnarray}
 {The symbol $U$ in the superscript indicates the fact that the ergotropy of a battery is the extractable energy from the battery using unitary operations.} 
 Consider the initial state of the auxiliary to be an arbitrary single-qubit state given by
    \begin{equation}\label{auxi}
        \rho_A=\frac{1}{2}(I_2+\vec{r}.\Vec{\sigma}),
    \end{equation}    \label{8}
    where $\Vec{r}=(r\sin{\theta_1}\cos{\phi_1},r\sin{\theta_1}\sin{\phi_1},r\cos{\theta_1})$ and $\Vec{\sigma}=(\sigma^x,\sigma^y,\sigma^z).$ Here $(r,\theta_1,\phi_1)$ are the spherical polar coordinates of the point which represents the state, $\rho_A$, in the Bloch sphere.
    %Therefore 
    %   \begin{equation}
    %    \rho_A=\frac{1}{2}(I+r\cos{\theta_1}\cos{\phi_1}\sigma_x+r\sin{\theta_1}\sin{\phi_1}\sigma_y+r\cos{\theta_1}\sigma_z).\nonumber
    %\end{equation}
          % 	\begin{figure}
		%\centering
	%\includegraphics[scale=0.22]{sep.pdf}
	%	\caption{Comparison between unitary- and measurement-based methods for energy extraction. We depict the behavior of $W^U(\rho_B,H_B)$ and $W^M_S(\rho_B,H_B)$ (shown along the vertical axis) with respect to the battery's initial parameter, $k$ (presented along horizontal axis). The green curve represents extractable energy using the probabilistic method, i.e, $W^M_S(\rho_B,H_B)$ whereas the blue curve describes energy extraction through unitary operation, $W^U(\rho_B,H_B)$. The horizontal axis is dimensionless, while the vertical axis is in the energy unit, $h$. The inset depicts the behavior of $W^M_S(\rho_B,H_B)-W^U(\rho_B,H_B)$, plotted along the vertical axis, with respect to $k$, presented along the horizontal axis. The vertical axis of the inset is again in the units of $h$ which has the energy unit and the horizontal axis is again dimensionless.}
	%	\label{fig:first}
	%\end{figure}
%Here, the orientation of the vector, $\Vec{r}$, in the Block sphere is defined by $\theta$ and $\phi$, and $r$ is the vector's modulus. 

According to our protocol, the system evolves under the action of the unitary, $U_H(t)$ [see Eq. \eqref{4eq3}], after which we perform the measurement on the auxiliary in a basis, $\{\kett{\psi},\kett{\psi_\perp}\}$. We obtain the distillable energy by maximizing over the parameters defining the initial auxiliary state, $(r,\theta_1,\phi_1)$, the total time of unitary evolution, $t$, as well as the parameters of the measurement basis, $(\theta,\phi)$. The optimization is carried out via Haar uniform generation of the relevant parameters, maintaining convergence up to the second decimal point. To emphasize the fact that, in this case, the initial battery-auxiliary state is considered to be separable, we denote the %probabilistically extractable
   {distillable} energy, $W^D(\rho_B,H_B)$, by $W^D_{S}(\rho_B,H_B)$.  In Fig.~\ref{non-linearity-graphx} (a), we plot the ergotropy, $W^U(\rho_B,H_B)$, and distillable energy, $W^{D}_S(\rho_B,H_B)$, along the vertical axis and $k$, whereas the parameter defining the initial state of $B$, is provided along the horizontal axis. The maximum energy obtained through unitary operation ($W^U$) and via the {measurement-based} process ($W^{D}_S$) are represented by blue and pink dots, respectively. It can be seen from Fig.~\ref{non-linearity-graphx} (a) that both the extractable energies, $W^U$ and $W_S^{D}$, are non-decreasing with $k$ though the behaviors of the functions are noticeably different for $k<0$ and $k>0$. Precisely $W^U=0$ for all $k<0$, whereas it is a linearly increasing function of $k$ for $k>0$.
  {The quantity $W_S^{D}$ increases linearly as $k$ varies from $-1$ to $1$, and remains greater than $W^U$ for all $k \neq -1, 1$. Moreover, the growth rate of maximum probabilistically extractable energy for initially product states ($W_S^{D}$) exceeds that of $W^U$ for every value of $k$, indicating that the distillable energy surpasses the ergotropy, except in the two limiting cases where the battery is initially prepared in $\kett{0}$ or $\kett{1}$.}
   %irrespective of whether the initial state is entangled or separable. In addition, initial entangled states yield greater distillable energy than separable states, except in the limiting cases where the battery is initially prepared in $\kett{0}$ or $\kett{1}$.}
   {We also present the behavior of $W_S^{M}$ as a function of $k$, and compare it with $W^U$ in Fig.~\ref{non-linearity-graphx} (b). From the figure, it is visible that at comparatively smaller values of $k$ where $W^U$ is zero, $W_S^M$ increases slowly with $k$. But just after crossing the threshold point, $k=0$, after which $W^U$ becomes non-zero, $W^M_S$ starts to rapidly increase with $k$.}
   This behavior is intuitively satisfactory, since for $k<0$, the population of the ground state of the initial battery state is larger than the same of the excited state, so that the initial energy of $B$ is rather small, and therefore it is more difficult to extract energy from such a state.
However, this does not explain the change in speed of the growth with respect to $k$ at $k=0$. {The significant difference between $W_S^{M}$ and $W^U$ for almost all values of $k$ indicates that even if a single suitable measurement outcome is chosen, that can still provide an advantage over energy extraction through the application of unitary on the battery.}

It is interesting to notice that though the extractable energy from a maximally mixed state, ${I}_2/2$, through unitary evolution is always zero independent of the Hamiltonian, {both the distillable energy and} the maximum probabilistically extractable energy from it is non-zero.
%for $k=0$. 
{To analyze the difference between the extractable energy using measurement-based and unitary-based methods in more detail, in Figs.~\ref{f3} (a) and \ref{f3} (b), we plot the differences, $W^{D}_S-W^{U}$ and $W^{M}_S-W^{U}$, respectively. From Fig.~\ref{f3} (a), it can be seen that within the range $-1<k<0$, the difference $W^{D}_S-W^{U}$ increases fast with an increase in $k$. This is because in this region, $W^{U}=0$. So $W^{D}_S-W^{U}$ is just the same as $W^{D}_S$. However, when $k$ is increased further, $W^{U}$ starts to increase with $k$ and $W^{D}_S$ cannot compete with the speed of increase in $W^{U}$. Hence, in the region $0\leq k\leq 1$, $W^{D}_S-W^{U}$ decreases with increasing $k$.
Consequently, we conclude that 
%probabilistic energy extraction, which is based on the best measurement on the best auxiliary system and at the best moment, 
extraction of distillable energy performs better than the best unitarily obtained energy. We also find a similar trend in the maximum probabilistically extractable energy, which is apparent from Fig.~\ref{f3} (b) where we plot the difference between $W^{M}_S$ and $W^{U}$.} 
{Since unitary operations preserve the eigenvalues, starting from a specific state, one cannot, in general, reach the ground state or any arbitrary state with lower energy that would allow for greater energy extraction, unless the final state has the same eigenvalues as the initial state.
However, in the measurement-based protocol presented here, a global unitary is involved, which allows the transfer of entropy from the battery to an auxiliary system, and vice versa. Therefore, the battery’s eigenvalues no longer remain fixed. This is the reason behind witnessing greater energy extraction, compared to the extractable energy by applying unitaries only on the battery.}
%\textcolor{blue}{Let us now discuss the optimal parameters which provide the 
%probabilistically extractable 
%distillable energy, $W^{D}_S$, when $B$ and $A$ are initially separable, and the time required to make $B$ and $A$ entangled enough so that $W^{D}_{S}$ amount of probabilistic extraction of energy is possible.}\textcolor{yellow}{eta vlo lgche ki sunte?}
{In Table~\ref{tab:updated_params} of Appendix~\ref{appA}, we present one set of optimal parameters representing measurement basis, time of evolution and initial auxiliary's state, for different initial states of $B$ defined through $k$. However, it should be noted that there are a large number of measurement settings, auxiliary states, and evolution times, using which we can squeeze out the %probabilistically extractable 
{distillable} energy from a given initial state of the battery. 
 For instance, for a given initial state, }{$\rho_B=\begin{bmatrix}
			0.7 & 0 \\
			0 & 0.3 \\
		\end{bmatrix}$, we show two choice of optimal parameter settings in Appendix~\ref{appA} both choices of parameter can distill the maximum possible energy from the battery using the measurement-based method even when the time of evolution is as small as 0.01$\frac{\hbar}{h}$.}
        %Here we have given a example of a choice of optimal measurement, optimal auxiliary and interaction time.

%In the next subsection, we show that if the joint state of the battery and auxiliary is entangled and pure, the entire energy of the battery can be \textcolor{blue}{distilled} using the measurement-based method, leaving the battery in its ground state, which is pure. 

\subsubsection{Comparing the time required for measurement- and unitary-based energy extraction methods}
We showed the energy extracted using measurement is much more than the maximum energy extractable through the application of unitaries on the battery. To analyze the efficiencies of the two mentioned methods more deeply, next we compare the time required to implement them. We know that the performance of measurements is instantaneous. But the measurement-based method also involves the application of a global unitary which is acted by switching on an interaction between the battery and the auxiliary for a fixed amount of time, $t$. To extract energy from the battery by directly applying unitary on it, one may also need to switch on a field for a fixed amount of time, say $t_U$. We want to compare $ t$ and $t_U$ which provides the maximum extractable energies.
In this regard, we consider a fixed initial state of the battery, i.e.,
%. Here the initial states are 
$\rho_B=\begin{bmatrix}
			\frac{1+k}{2} & 0 \\
			0 &\frac{1-k}{2} \\
		\end{bmatrix}$. We take $k>0$, otherwise $\rho_B$ will be passive. 
In the measurement-based method, {i.e. the one where distillable energy is calculated,} the interaction between the auxiliary and the battery is governed by $H_{BA}$, defined in Eq. \eqref{4eq5}. It is clear from Table \ref{tab:updated_params}, for all $k$, the time of interaction, $t$, is always less than 1 $\frac{\hbar}{h}$.% Hence not only the probabilistically extractable energy is much greater than the ergotropy, but also the time, $t$, required to extract that energy is much lesser than the time, $t_U$, needed in the unitary-based method. We would like to note here the trace norm of the battery's local Hamiltonian, $H_B$, involved in $H_{BA}$, is $2h$. 

Let us now scrutinize the unitary-based energy extraction method. The passive state having the same eigenvalues as $\rho_B$ is
       $\sigma_{\rho_B}=\begin{bmatrix}
			\frac{1-k}{2} & 0 \\
			0 &\frac{1+k}{2} \\
		\end{bmatrix}$. 
Hence, to extract the maximum possible amount of energy from $\rho_B$, one needs to apply a unitary that transforms $\rho_B$ to $\sigma_{\rho_B}$. We know $\sigma^x\rho_B\sigma^x=\sigma_{\rho_B}$. To create $\sigma^x$, one may switch on a field described by the Hamiltonian $H_B'=h(I_2-\sigma_x)$ for the $t_U=\pi/2$ $\frac{\hbar}{h}$ amount of time. Here we consider $H_B'$ in such a way that it has the same trace norm as $H_B$, that is, $2h$, to make the comparison between $t$ and $t_U$ meaningful. 

Hence we see $t$ is always much less than $t_U$ for all values of $k$. We would like to add here that in Table \ref{tab:updated_params} we present one set of optimal parameters. But as we mentioned before, the set of optimal parameters is not unique. Therefore, $t$ can even be reduced, keeping the %probabilistically extractable 
{distillable} energy the same.

\subsection{Case II: Entangled initial battery-auxiliary state}\label{2xb}
\label{b}
{In the previous subsection, we witnessed the advantage of performing projective measurement on the auxiliary system, which was attached to the battery, in energy extraction. In that case, the entire battery-auxiliary system was initially considered to be in a separable state.} In this section, instead of restricting ourselves to separable initial states, we consider non-zero entanglement to be present in the initial battery-auxiliary state. We wish to examine the impact of this initial entanglement on energy extraction.

Let the initial state of the battery and auxiliary, $\ket{\psi_{BA}}$, be of the following form, 
\begin{equation}
    \ket{\psi_{BA}}=\sqrt{\frac{1+k}{2}}\ket{0}\ket{\phi}+\sqrt{\frac{1-k}{2}}\ket{1}\ket{\phi_{\perp}}.
    \end{equation}
    Where $\ket{\phi}$ and $\ket{\phi_{\perp}}$ are arbitrary orthogonal pure states in $\mathcal{H}_A$. 
    Let $\rho_{BA}=\ket{\psi_{BA}}\bra{\psi_{BA}}$.
    The particular form of $\kett{\psi_{BA}}$ is dictated by the fact that if we trace out the auxiliary part, the initial state of the battery reduces to  
    \begin{equation}
    \Tr_A{(\rho_{BA})}=\rho_B=\begin{bmatrix}
			\frac{1+k}{2} & 0 \\
			0 &\frac{1-k}{2} \\
		\end{bmatrix}\label{4eq2},
    \end{equation}
which is the same initial battery state considered in the previous sub-section. This similarity in the initial states considered in Secs. \ref{a} and \ref{b} will allow us to compare the situation of having entanglement as a resource with the one of not having it.

The general expressions of $\ket{\phi}$ and $\ket{\phi_{\perp}}$ can be considered to be
        \begin{eqnarray}\label{ent_11}
        \ket{\phi}&=&\cos(\frac{\theta_2}{2})\ket{0}+\exp(-i\phi_2)\sin(\frac{\theta_2}{2})\ket{1}\nonumber\\\text{ and}
        \ket{\phi_{\perp}}&=&\sin(\frac{\theta_2}{2})\ket{0}-\exp(-i\phi_2)\cos(\frac{\theta_2}{2})\ket{1}.
    \end{eqnarray}
\begin{figure}[htbp]
    \subfigure[]{\includegraphics[width=0.23\textwidth]{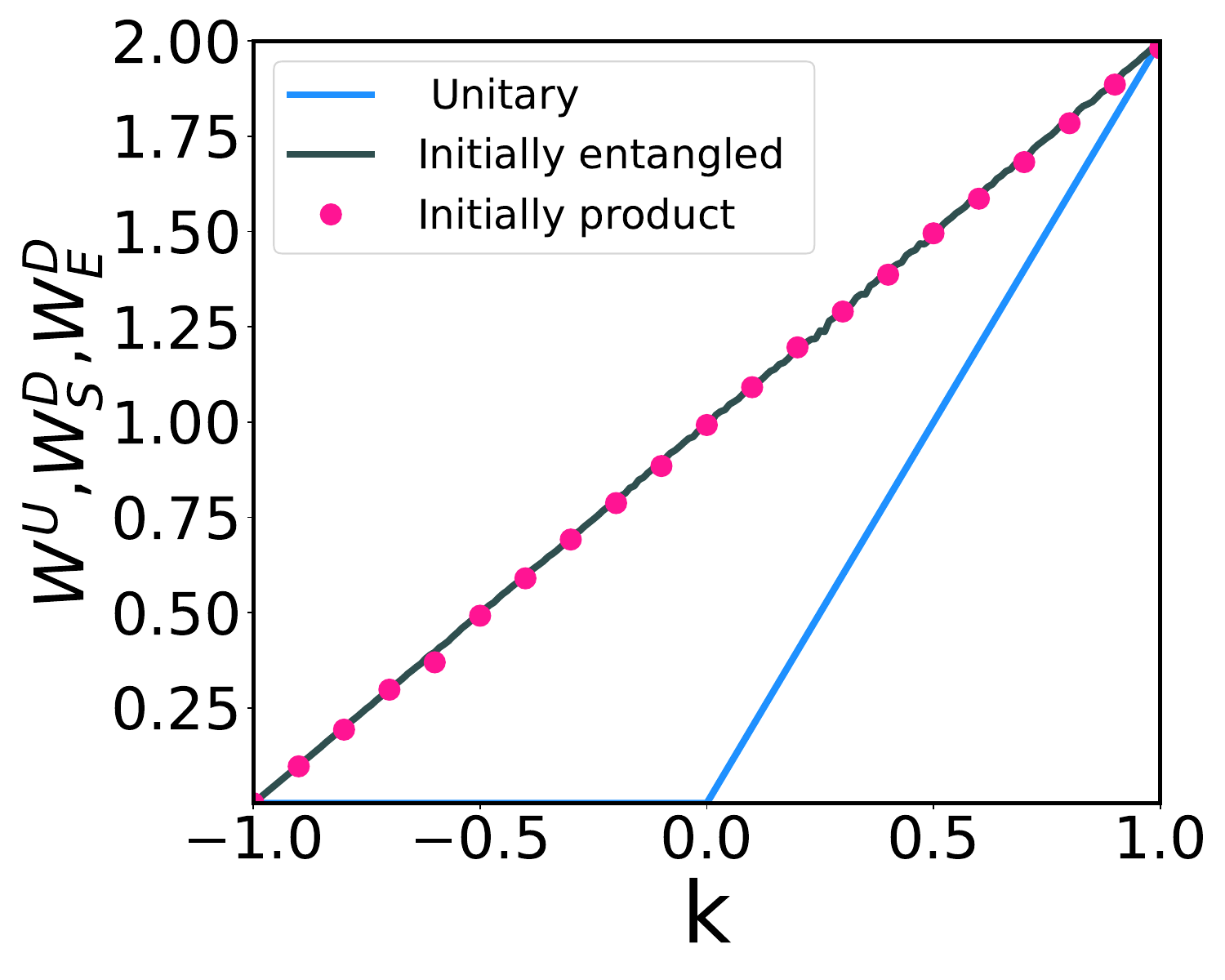}} 
    \label{popx}
    \subfigure[]{\includegraphics[width=0.23\textwidth]{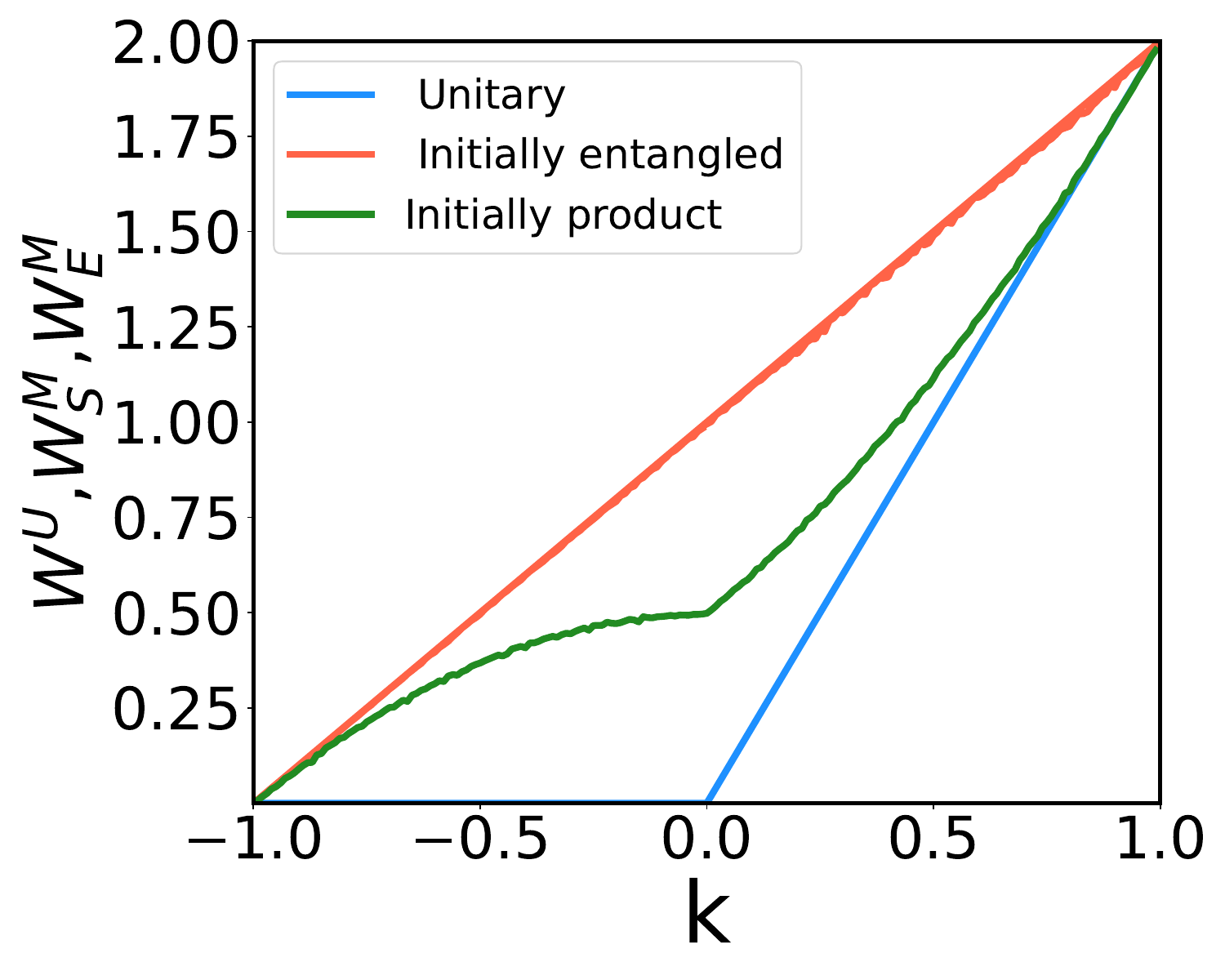}} 
    \label{1x1b}
    \caption{{Dependence of extractable energies on the initial state of the battery. In Fig.~\ref{non-linearity-graphx} (a), we plot $W^U(\rho_B,H_B)$ (blue line), $W^{MD}_S(\rho_B,H_B)$ (pink dot), and $W^{MD}_E(\rho_B,H_B)$ (dark slate gray) along the vertical axis as a function of $k$ presented on the horizontal axis. Here $W^{MD}_S(\rho_B,H_B)$ and $W^{MD}_E(\rho_B,H_B)$ are the terms that denote the distillable energy when the battery and auxiliary are initially product and initially entangled states. On the other hand, in Fig.~\ref{non-linearity-graphx} (b), $W^U(\rho_B,H_B)$ (blue line), $W^{M}_S(\rho_B,H_B)$ (green line), and $W^{M}_S(\rho_B,H_B)$ (red line) are plotted along the vertical axis, and the horizontal axis represents the initial battery state's parameter $k$. Here the terms $W^{M}_S(\rho_B,H_B)$ and $W^{M}_S(\rho_B,H_B)$ denote the maximum probabilistically extractable energy when the battery and auxiliary are initially product and initially entangled states.} }
    \label{non-linearity-graphx}
\end{figure}

\begin{figure*}[!htbp]
    \subfigure[]{\includegraphics[width=0.28\textwidth]{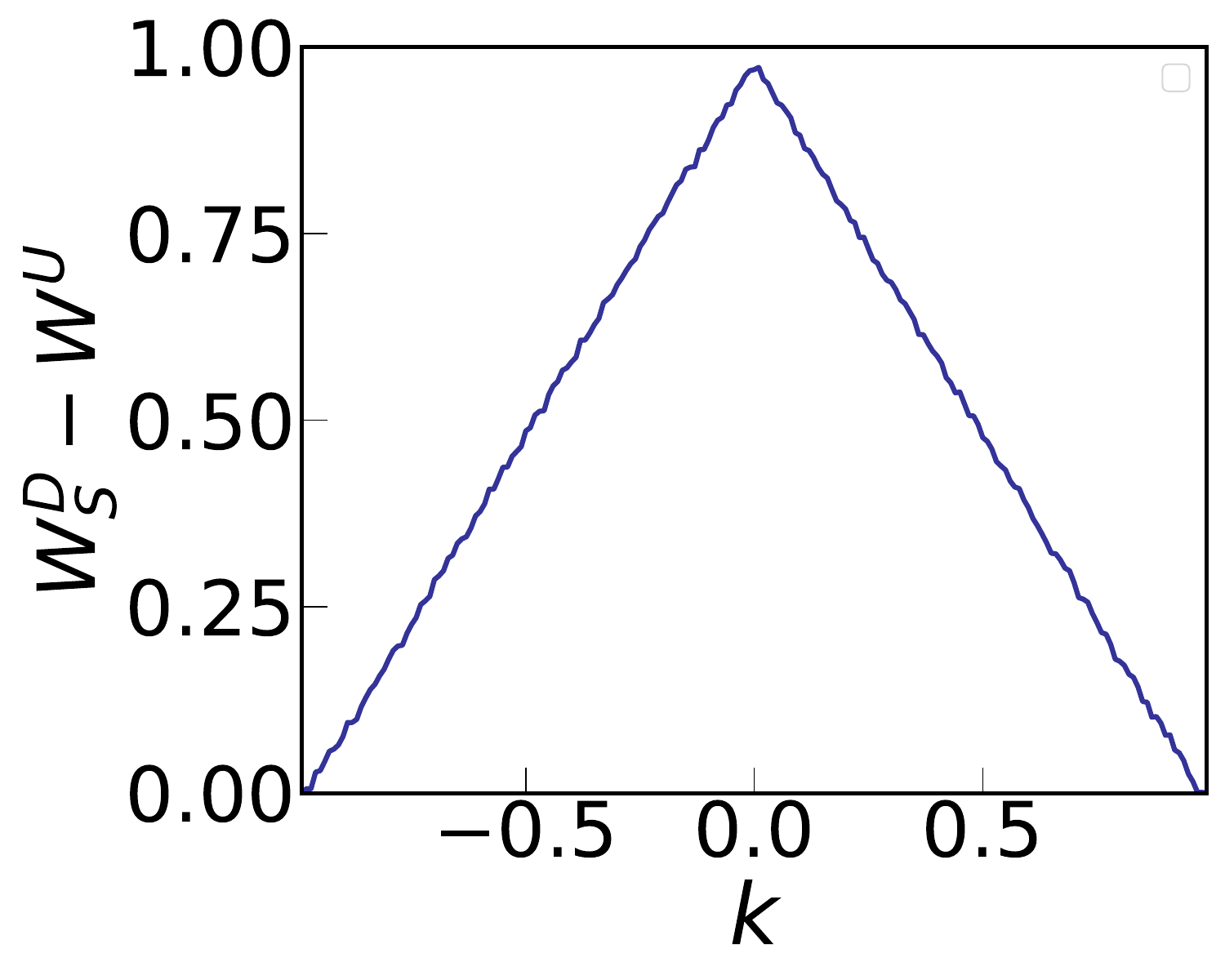}} 
         \label{x1}
             \hspace{0.32cm}
    \subfigure[]
{\includegraphics[width=0.28\textwidth]{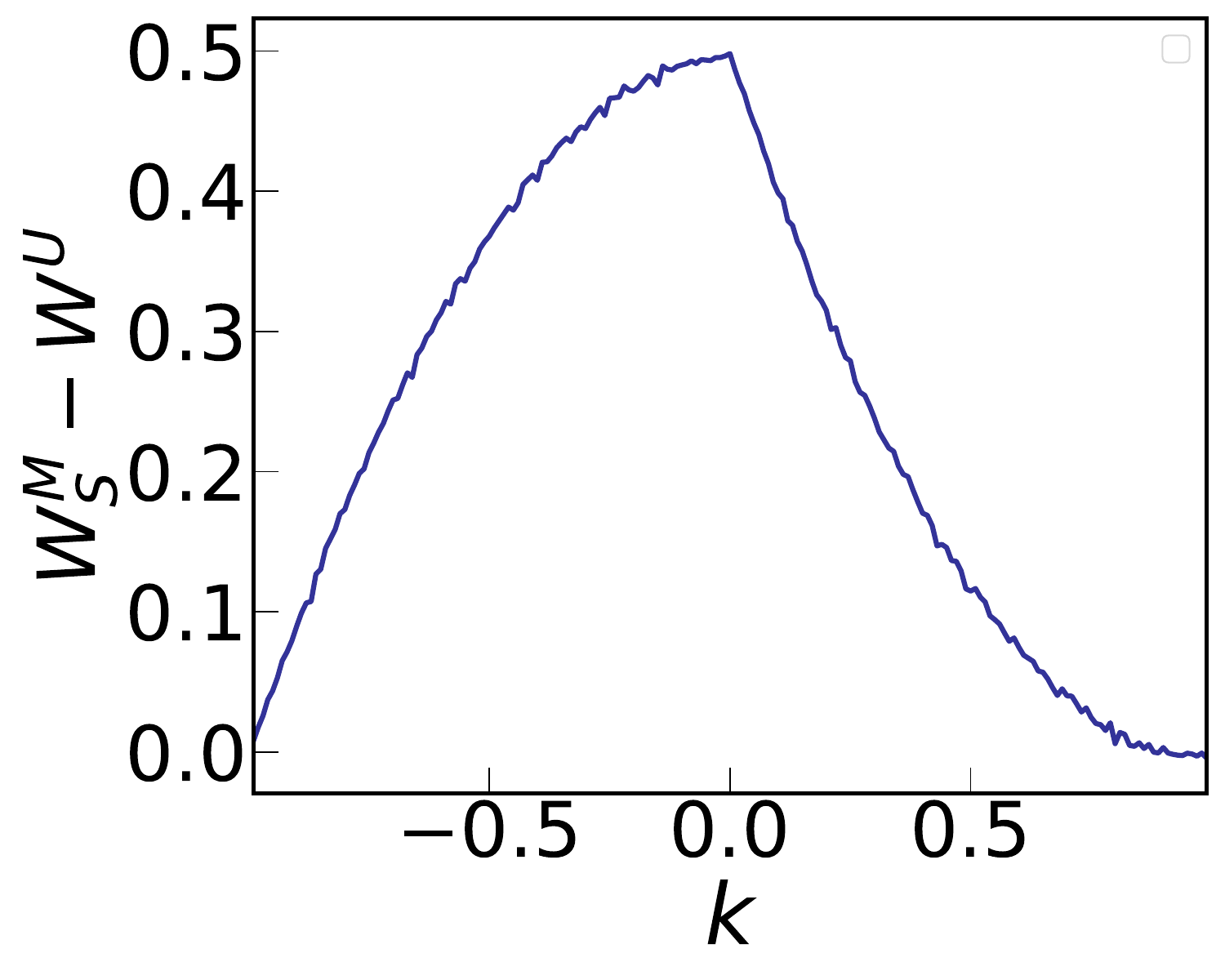}} 
         \label{x1}
         \hspace{0.31cm}
             \subfigure[]{\includegraphics[width=0.30\textwidth]{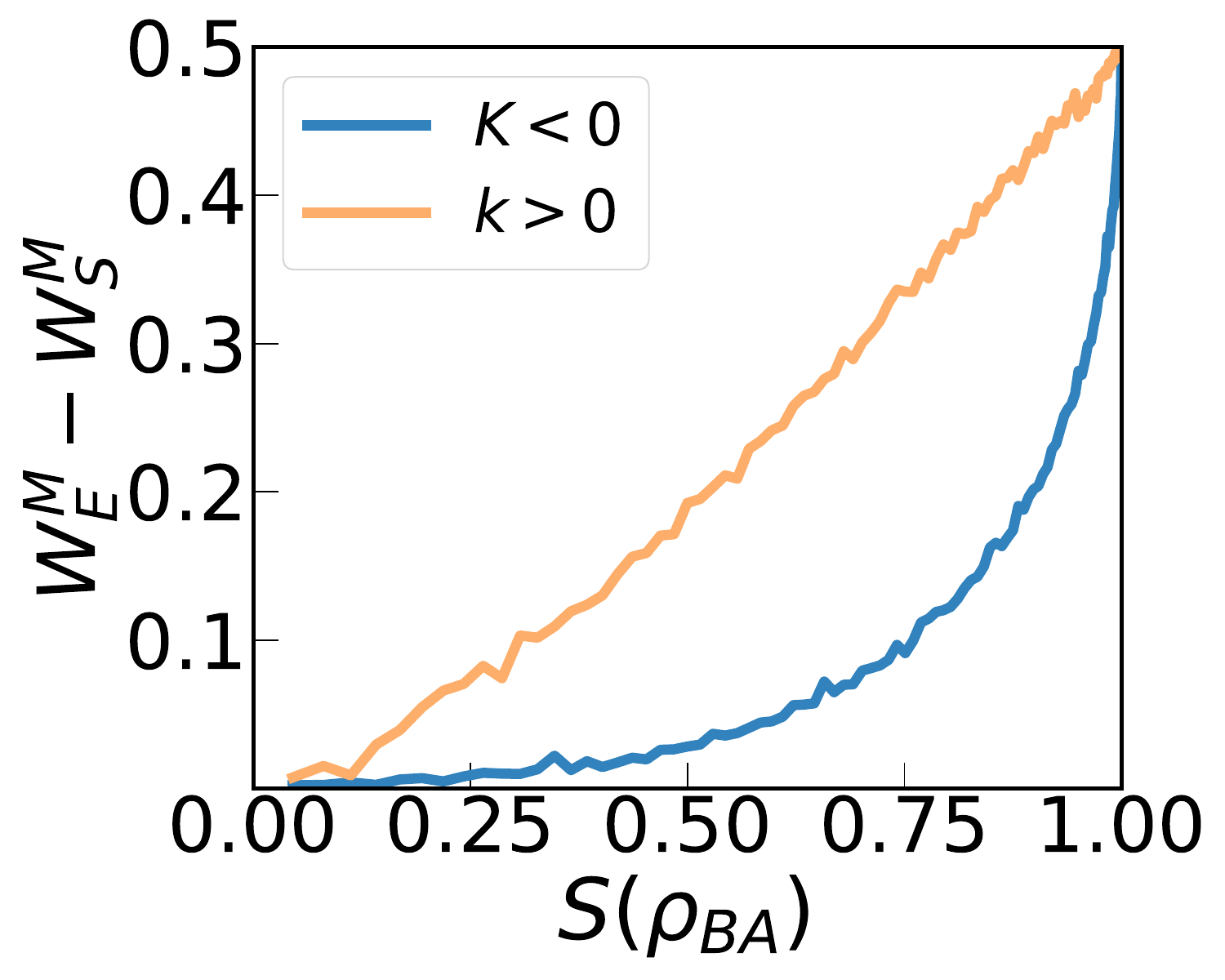}} 
         \label{x1}

\caption{{$W^{D}_S(\rho_B,H_B)-W^U(\rho_B,H_B) $ vs. $k$, $W^{M}_S(\rho_B,H_B)-W^U(\rho_B,H_B) $ vs. $k$ and $W^{M}_E(\rho_B,H_B)-W^{M}_S(\rho_B,H_B) $ Vs $S(\rho_{BA}) $ plots are depicted in Fig.~\ref{f3} (a), Fig.~\ref{f3} (b), and Fig.~\ref{f3} (c). Fig.~\ref{f3} (a) and Fig.~\ref{f3} (b) $W^{D}_S(\rho_B,H_B)-W^U(\rho_B,H_B)$ and $W^{M}_S(\rho_B,H_B)-W^U(\rho_B,H_B)$, plotted along the vertical axis, with respect to $k$, presented along the horizontal axis. The vertical axis is in the units of $h$, which has the energy unit, and the horizontal axis is again dimensionless. In Fig.~\ref{f3} (c), we depict $W^{M}_E(\rho_B,H_B)-W^{M}_S(\rho_B,H_B)$ with respect to the entanglement content of the initial state of the battery and the auxiliary. Here the orange and blue curves represent the parameter regions $k>0$ and $k<0$, respectively. The entanglement is quantified using entanglement entropy. The horizontal axis of the main plot is dimensionless, whereas the vertical axes of both of the plots are in the units of $h$, that is, in energy units. The horizontal axis of the inset, i.e., the entanglement entropy, is plotted in units of ebits.}}
        \label{f3}
\end{figure*}
We repeat the same measurement-based protocol considering $\rho_{BA}=\kett{\psi_{BA}}\braa{\psi_{BA}}$ as the initial state. We optimize over $\theta_2$ and $\phi_2$ which describes the initial state of the auxiliary, the time $t$ at which the measurement has been done, and the measurement-projectors, in particular $\theta$ and $\phi$, to obtain the 
{distillable energy, $W^{D}(\rho_B,H_B)$}.
%probabilistically extractable energy, $W_{\max}^M(\rho_B,H_B)$. 
Since here we considered the joint initial state describing the battery and the auxiliary qubits to be entangled, {we denote the distillable energy in this process by $W_{E}^{D}$}.

{To compare the nature of $W_{E}^{D}$ with the scenario where the joint initial state of $B$ and $A$ was separable, we plot $W_{S}^{D}$ with pink dots in the same figure where $W_{E}^{D}$ are denoted with dark slate gray curves, i.e., Fig.~\ref{non-linearity-graphx} (a). 
 One can notice from the figure that both $W^{D}_S$ and $W^{D}_E$ are equal for all the values of the initial state parameter $k$ and higher than the value of $W^{U}$ for almost all values of $k$, except $k=\pm1$. {For completeness, we also analyze the probabilistically extractable energy, $W_E^M$, considering initially entangled battery-auxiliary state as taken for distillable energy. We find $W_E^M$ to be identical to $W_S^D$ and $W_E^D$ for every considered value of $k$. We plot $W_E^M$ along with $W_S^M$ and $W^U$ with respect to $k$ in Fig.~\ref{non-linearity-graphx} (b).}
 Therefore, we can say that the initial entanglement between the battery and the auxiliary does not provide any benefit in energy distillation. However, the presence of entanglement provides  benefit in probabilistically extractable energy. It is to be noted that the maximum probabilistically extractable energy for the initially entangled state and the distillable energy for both the initially entangled and product states provide the complete extraction, which is equal to the energy difference between the energy of the initial battery's state and the ground state of the battery's Hamiltonian.
 To explore how this advantage depends on the amount of entanglement of $\kett{\psi_{BA}}$, in Fig.~\ref{f3} (c),  the difference between $W^{M}_E$ and $W^{M}_S$ is plotted with respect to the entanglement entropy \cite{entropy__1} of $\rho_{BA}$, i.e. the binary entropy of $\rho_B$ given by $S(\rho_{B})=-\mathrm{tr}{\rho_B \log_2 \rho_B}$.}
%, where 
%$H(\cdot)$ 
%$S(\rho_B)$  denotes the binary entropy function of the state $\rho$.
%represents the entropy function. 
%At the same time t
The binary entropy of any qubit state $\rho$ having eigenvalues $p$ and $1-p$ is given by $S(\rho)=-(1-p) \log_2 (1-p)-p \log_2 p$. 
%Here, $S(\rho)$  denotes the binary entropy of the state $\rho$}.  
%Let us examine how exactly the initial entanglement enhances the energy extraction. 
One can notice that the entanglement entropy of the initial state for $k=x$ is equal to the case where $k=-x$ for all values of $x$ between 0 and 1. Since we are varying $k$ from -1 to 1, for each negative value of $k$ there will be a corresponding positive value of $k$ for which the state will have the same entanglement as for the negative of $k$. Thus, we have two curves, one for $k<0$ (cyan) and one corresponding to $k>0$ 
%(purple)
(orange). Observing both the curves, we find that not only is the presence of entanglement beneficial, but also the advantage increases with the entanglement. Clearly, the behavior of this increment with entanglement is different for $k>0$ and $k<0$. We speculate that the small fluctuations seen in the plots of the figure are occurring because of the limitation in numerical precision. 

{We therefore find that, for a given marginal battery state, the presence of initial entanglement enhances energy extraction in the measurement-based protocol. One needs to keep in mind here that ergotropy, i.e., the maximum energy extractable from a battery by applying a unitary on that battery, is solely dependent on the battery's initial state. In particular, for a given quantum battery, the ergotropy depends on its initial state, and once the initial state is given, one can uniquely find its ergotropy even if it is entangled with the rest of the world and the entanglement is unknown. As a result, in Figs.~\ref{non-linearity-graphx} (a) and~\ref{non-linearity-graphx} (b), the ergotropy (blue curve), $W^U$, remains the same, independent of the fact if the battery is entangled with the auxiliary or not, because in both cases, the marginal state of the battery is taken to be the same, which implies that the eigenspectrum of the battery is also the same in both situations.}

\begin{figure}[b]
    \subfigure[]{\includegraphics[width=0.43\textwidth]{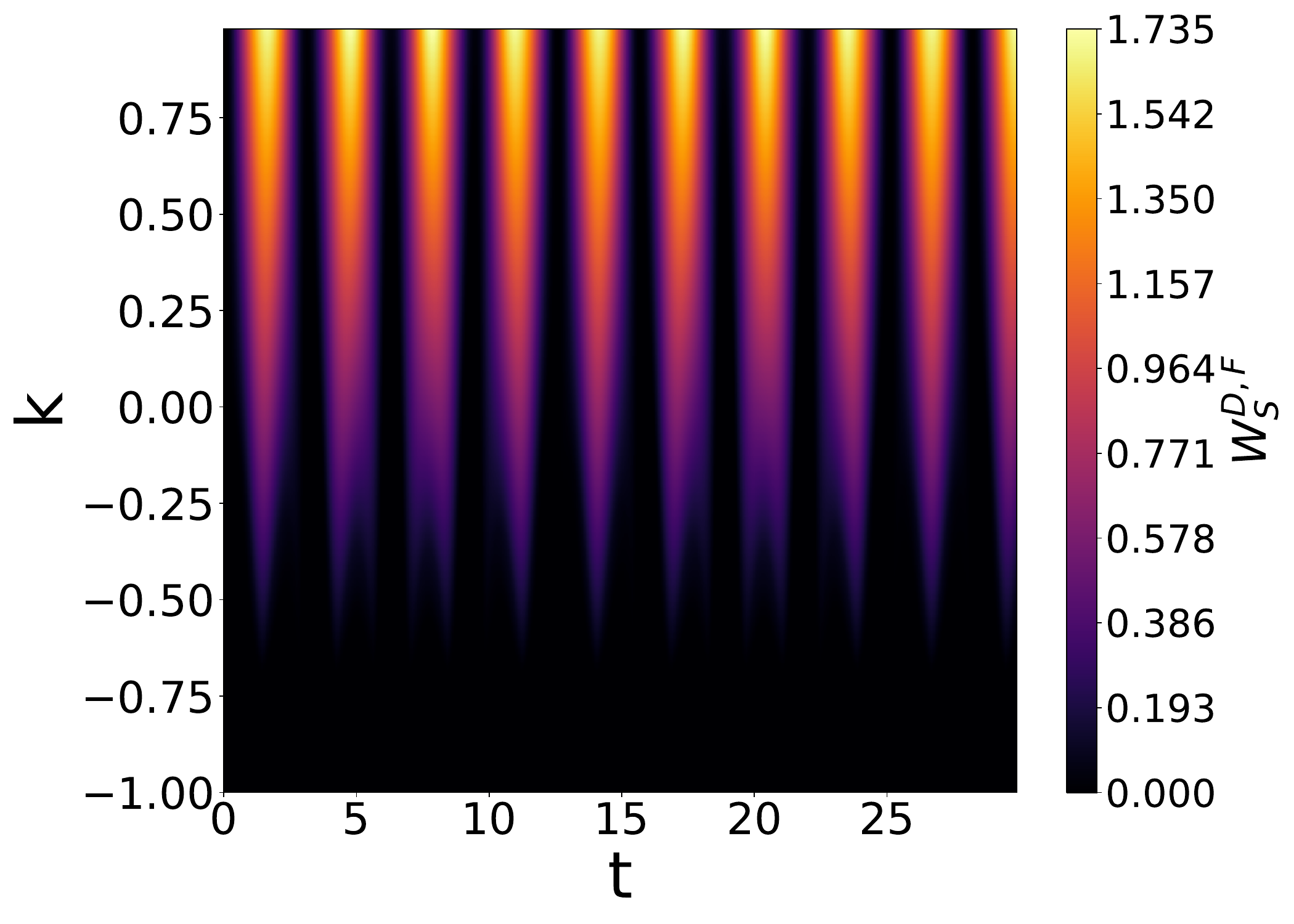}}
        \subfigure[]{\includegraphics[width=0.43\textwidth]{ 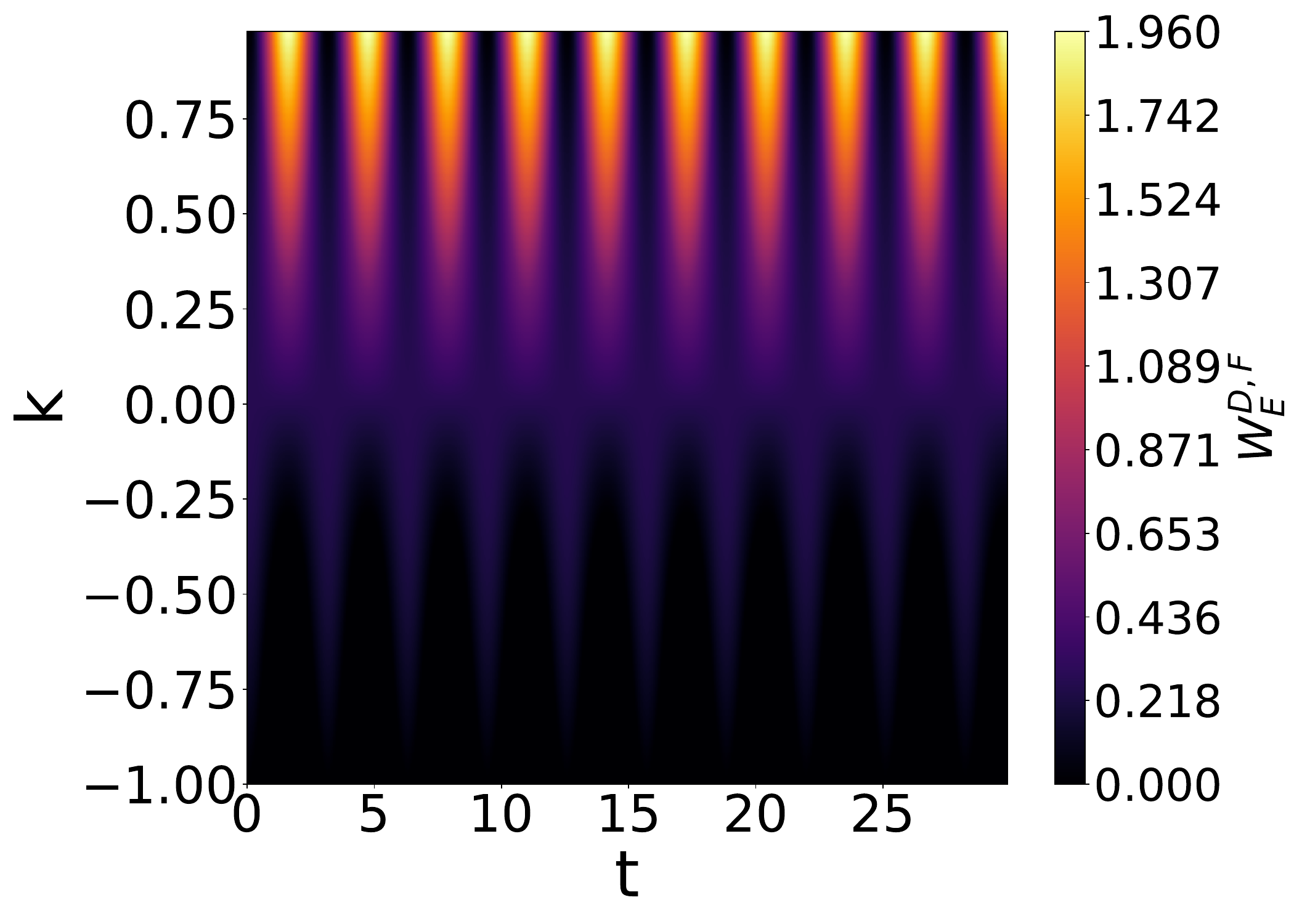}}
\caption{{In Fig.~\ref{4f} (a), we plot distilled energy $W^{D,F}_S$ in a 2D projection graph for a fixed measurement setting and fixed auxiliary system (given in the last part of subsection~\ref{2xb}) with the initial state parameter $k$ along the vertical axis and time $t$ of interaction through Hamiltonian $H_{BA}$ along the horizontal axis when the battery and auxiliary are initially product. On the other hand, Fig.~\ref{4f} (b) describes the same numerical studies for initially entangled states where the horizontal and vertical axes have the same notation as Fig.~\ref{4f} (a). The distilled energy for a fixed measurement setting and fixed auxiliary system for initially entangled states is denoted as $W^{D,F}_E$. The values of $W^{D,F}_S$ and $W^{D,F}_E$ for a given $t$ and $k$ values are denoted by color bars of respective plots.}}
        \label{4f}
\end{figure}

{ 
Next, we investigate how the extractable energy evolves with time $t$ for a fixed auxiliary system and measurement setting, considering both initially product and initially entangled states. For the initially product states, the auxiliary system (described in Eq.~\ref{auxi}) parameters are chosen as $r = 0.7$, $\theta_1 = \pi$, and $\phi_1 = 2\pi$, and for the initially entangled state, the auxiliary system's (described in Eq.~\ref{ent_11}) parameters are $\theta_1 = \pi$ and $\phi_1 = 2\pi$. The measurement setting is fixed at $\theta = 1$ and $\phi = 2\pi$ for both cases.
We find that the distilled energy for the fixed measurement setting, fixed auxiliary, exhibits an oscillatory behavior over time for both the initial product and the entangled state, as depicted in Fig.~\ref{4f} (a) and Fig.~\ref{4f} (b), respectively. 
Here, the distilled energy for a fixed measurement and fixed auxiliary system for both the initial product and entangled state is plotted in a 2D projection graph to their respective subplots, where the initial state parameter $k$ is along the vertical axis, and the time $t$ of interaction is along the horizontal axis. $W^{D,F}_S$ and $W^{D,F}_E$ denote the distilled energy for initially product and entangled states, respectively, with a fixed measurement setting and fixed auxiliary system. The color bars in the corresponding subplots are used to indicate the values of $W^{D,F}_S$ and $W^{D,F}_E$ for particular $t$ and $k$ values.  For distilled energy, the typical time period of oscillation for initially product and entangled states at $k = 0.2$ is found to be $3.01$ and $3.167$ in the unit of $\frac{\hbar}{h}$.
At the same time 
the same values of parameters are considered in Fig.~\ref{1a} (a) and Fig.~\ref{1a} (b) of the Appendix~\ref{appA}.
This periodicity with time $t$ is also captured for probabilistically extractable energy. The probabilistically extractable energy for the initially product and entangled state is denoted as $W^{M,F}_{S}$ and $W^{M,F}_E$. On the other hand, for probabilistically extractable energy, the typical time period of oscillation for initially product and entangled states at $k = 0.2$ is found to be $3.08$ and $3.136$  in the unit of $\frac{\hbar}{h}$.}
{\section{Comparison between daemonic ergotropy and distillable energy}\label{3x1}}
{
%Along with that, 
In this section, we compare the daemonic ergotropy, introduced in~\cite{demon_1}, with the distillable energy. 
%
% In general, daemonic ergotropy is defined as 
% \begin{equation}
%     W_D=\Tr[\rho_B\sigma_z]-\sum_i\min_{u_i}\Tr[\rho_i^{D}\tilde{\sigma}^z_i],
% \end{equation}}  
% \textcolor{blue}{where $\rho_i^{D}=\Tr_A[\rho_{BA}(\mathbbm{I}\otimes M_i)] $ and $\tilde{\sigma^z_i}=u_i\sigma^z_iu^{\dag}_i$. Here $u_i$ is a unitary operator corresponding to the $i^{\text{th}}$ measurement outcome acting on the local state of the battery.
% $M_i$ is the $i^{\text{th}}$ measurement operator as defined previously.
%
In general, the daemonic ergotropy is defined as  
\begin{equation}
    W_D = \Tr[\rho_B \sigma_z] - \sum_i \min_{u_i} \Tr[\rho_i^{D} \tilde{\sigma}^z_i],
\end{equation}
where $\rho_i^{D} = \Tr_A\!\left[\rho_{BA}(\mathbbm{I} \otimes M_i)\right]$ denotes the post-measurement state of the battery after tracing out the ancillary system, and $\tilde{\sigma}^z_i = u_i \sigma^z_i u_i^{\dagger}$ represents the Pauli-$z$ operator in the optimally rotated basis determined by the unitary $u_i$ corresponding to the $i^{\text{th}}$ measurement outcome. Here, $M_i$ is the $i^{\text{th}}$ measurement operator as defined previously.
%
% %Now, o
% Our initial state is $\ket{\psi_{BA}}=\sqrt{\frac{1+k}{2}}\ket{0}\ket{\phi}+\sqrt{\frac{1-k}{2}}\ket{1}\ket{\phi_{\perp}}. $ Now, if we consider 
% %our
% the measurement setting 
% %as 
% to be $M_1=\ket{\phi}\bra{\phi}$ and $M_2=\ket{\phi_{\perp}}\bra {\phi_{\perp}}$, then
% %. Now 
% measurement in the $M_1$ and $M_2$ basis will provide the outcomes $\sqrt{\frac{1+k}{2}}\ket{0}$ and $\sqrt{\frac{1-k}{2}}\ket{1}$ with probabilities $\frac{1+k}{2}$ and $\frac{1-k}{2}$, respectively. Therefore, the daemonic ergotropy is $W_D=k+1$. This means that the daemonic ergotropy provides maximum possible energy extraction, as $k+1$ is basically the energy difference between the state $\rho_B=\frac{1+k}{2}\ket{0}\bra{0}+\frac{1-k}{2}\ket{1}\bra{1}$ and the ground state $\ket{1}\bra{1}$. %But 
% However, if the composite system is initially in the general product state $\rho_{BA}=\rho_B\otimes \rho_A$, then local measurement on the auxiliary system has no effect on the state of the battery, and will be equal to the ergotropy. Therefore, although for initial entangled state, the daemonic ergotropy and distillable energy are the same, for initial product state, the
% %amount of 
% distilable energy is higher than the daemonic ergotropy.  
%
Consider the initial state  
\begin{equation}
    \ket{\psi_{BA}} = \sqrt{\frac{1+k}{2}}\,\ket{0}\ket{\phi} + \sqrt{\frac{1-k}{2}}\,\ket{1}\ket{\phi_{\perp}}.
\end{equation}
If we choose the measurement operators  
\begin{equation}
    M_1 = \ket{\phi}\bra{\phi}, \quad M_2 = \ket{\phi_{\perp}}\bra{\phi_{\perp}},
\end{equation}
then measurement in the $\{M_1, M_2\}$ basis projects the battery's state to  
$\ket{0}$ or $\ket{1}$ with probabilities $\frac{1+k}{2}$ and $\frac{1-k}{2}$, respectively.  
In this case, the daemonic ergotropy evaluates to $W_D = k+1$, which corresponds to the maximum possible energy extraction.  
Indeed, $k+1$ equals the energy difference between the state  
\begin{equation}
    \rho_B = \frac{1+k}{2}\ket{0}\bra{0} + \frac{1-k}{2}\ket{1}\bra{1}
\end{equation}
and the ground state, $\ket{1}\bra{1}$.  
However, if the composite system is initially in a product state  
$\rho_{BA} = \rho_B \otimes \rho_A$, then local measurements on the auxiliary system leave the state of the battery unchanged, and the daemonic ergotropy reduces to the standard ergotropy.  
Therefore, while for an initially entangled state the daemonic ergotropy coincides with the distillable energy and maximum probabilistic extractable energy, for an initial product state both the distillable energy and maximum probabilistic extractable energy exceeds the daemonic ergotropy.}

%Therefore,  

In our work, we compare three methods of energy extraction: unitarily extracted energy, energy extracted via measurement when the battery and auxiliary are uncorrelated, and energy extracted via measurement when the battery and auxiliary are correlated. In all three cases, the fully charged state corresponds to the excited state of the Hamiltonian, but the empty or passive state, defined as the state from which no energy can be extracted, differs across the protocols. If we take the initial battery-auxiliary state to be a product, both the distillable and maximum probabilistically extractable energies are non-zero for all battery states, except the ground state,

For both the measurement-based protocols, the ground state of the Hamiltonian serves as the empty state, whereas for the unitary-based extraction protocol, passive states depend on the mixedness of the initial state. Notably, when the initial state is a pure state, the ground state acts as the empty state for all three methods, highlighting the role of initial state properties .

We would like to emphasize here that the measurement-based protocol involves two steps: action of a global unitary on the joint state of the battery and the auxiliary, followed by a projective measurement on the auxiliary. We are investigating the effects of both steps together as a whole, and not separately. Therefore, the advantage of this method shown by comparing it to the unitary-based energy extraction process originates from the joint process, which involves measurement as well as the correlation between the battery and the auxiliary created through interaction. For example, if the auxiliary battery were not correlated, the measurement would not have any effect on the battery, resulting in zero-energy extraction. To understand the amount of advantage that arises from the measurement and not from the correlation between the $A$ and $B$, in the next part, we compare the maximum extractable energies in the measurement-based protocol and using the global unitary acting on the battery and the auxiliary, which is used in the measurement-based protocol to correlate $B$ and $A$.

{\section{Energy Extraction and Power Output via Auxiliary-Battery Interaction}\label{s4cx}}

Here we will first find the extractable energy using $U_H(t)$, then compare it with the %probabilistically extractable 
{distillable} energy.
We will consider two scenarios: one where the battery and auxiliary are initially products and the other where they are initially entangled.

%Rather of measuring the auxiliary system in this section, we just allow the battery and auxiliary to evolve globally before tracing out the auxiliary system. Next, we determine the maximum extractable energy for each of the two cases that is considered below.}

Let us begin with the first case where the initial state of the battery is $\rho_{B}=\frac{(1+k)}{2} \ket{0}\bra{0}+\frac{(1-k)}{2} \ket{1}\bra{1}$ and that of the auxiliary is given by $\rho_A=\frac{(I+\vec{r}.\vec{\sigma})}{2}$. Thus the initial state of the composite system is of the form $\rho_{BA}(0)=\rho_B \otimes \rho_A$.
%\textcolor{red}{In the first scenario, the system and auxiliary are initially products. On the other hand, in the next case, we consider the initial state of battery and auxiliary as entangled. Let us begin with the first case, where the initial state of the battery is $\rho_{B}=\frac{(1+k)}{2} \ket{0}\bra{0}+\frac{(1-k)}{2} \ket{1}\bra{1}$. On the other hand, the initial state of the auxiliary is given by $\rho_A=\frac{(I+\vec{r}.\vec{\sigma})}{2}$. Here, the initial state of the composite system is taken as a product, that is, $\rho_B \otimes \rho_A$.}
The joint state of the system and auxiliary undergoes a unitary evolution via the Hamiltonian $H_{BA}$, i.e., $U_H(t)=\exp(-i H_{BA} t)$. The evolved state, at a time $t$, is given by $\rho_{BA}(t)=U_H(t)\rho_B \otimes \rho_A U_H^{\dag}(t)$. The extractable energy in this process is defined as
\begin{eqnarray*}
    W_{S}^{U_H}&=&\Tr[\rho_{B} H_B]\\
    &&-\min_{t, \rho_A}\Tr\left[\Tr_A\left(U_H(t)\rho_B \otimes \rho_A U_H^{\dag}(t)\right) H_B\right].\nonumber
\end{eqnarray*}

%\textcolor{red}{After that, the joint state goes through a unitary evolution by the hamiltonian $H_{BA}$, that is, $U=\exp(-i H t)$. After the evolution of the initial joint state, we get the final initial state at a time $t$, given by $\rho^{t}_{BA}=U\rho_B \otimes \rho_A U^{\dag}$. After that, we trace out the system auxiliary and calculate the optimal extarctable energy, where the optimization is done over all possible auxiliary systems and time $t$. }

In the second case, a similar protocol is followed. The initial entangled state considered here is $\ket{\psi}_{BA}=\frac{(1+k)}{2}\ket{0}\ket{\phi}+\frac{(1-k)}{2}\ket{1}\ket{\phi_{\perp}}$, where $\ket{\phi_\perp}$ is an orthogonal state of $\ket{\phi}$. One can notice, the reduced state of $B$ is the same as in the previous case where $B$ and $A$ were products.
{%In the second case, the same thing is done for the initially entangled state, that is $\ket{\psi}_{BA}=\frac{(1+k)}{2}\ket{0}\ket{\psi}+\frac{(1-k)}{2}\ket{1}\ket{\psi_{\perp}}$. 
The joint state at time $t$ after unitary evolution 
%Here, at a time $t$ the evolve state 
is given by, $\ket{\psi}^t_{BA}=U_H(t)\ket{\psi}_{BA}$. 
   	\begin{figure}[b]
    \subfigure[]{\includegraphics[width=0.23\textwidth]{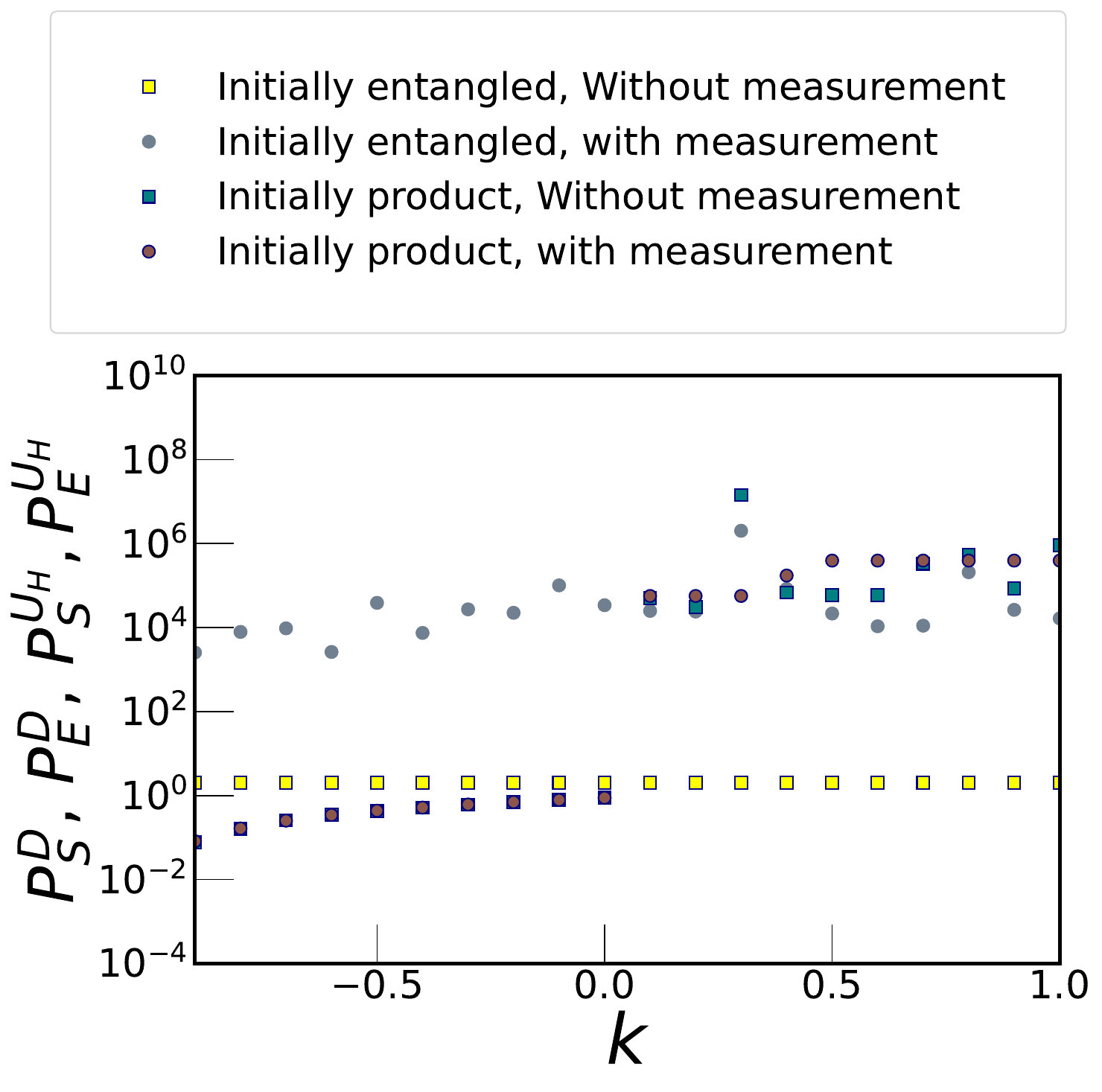}}
        \subfigure[]{\includegraphics[width=0.23\textwidth]{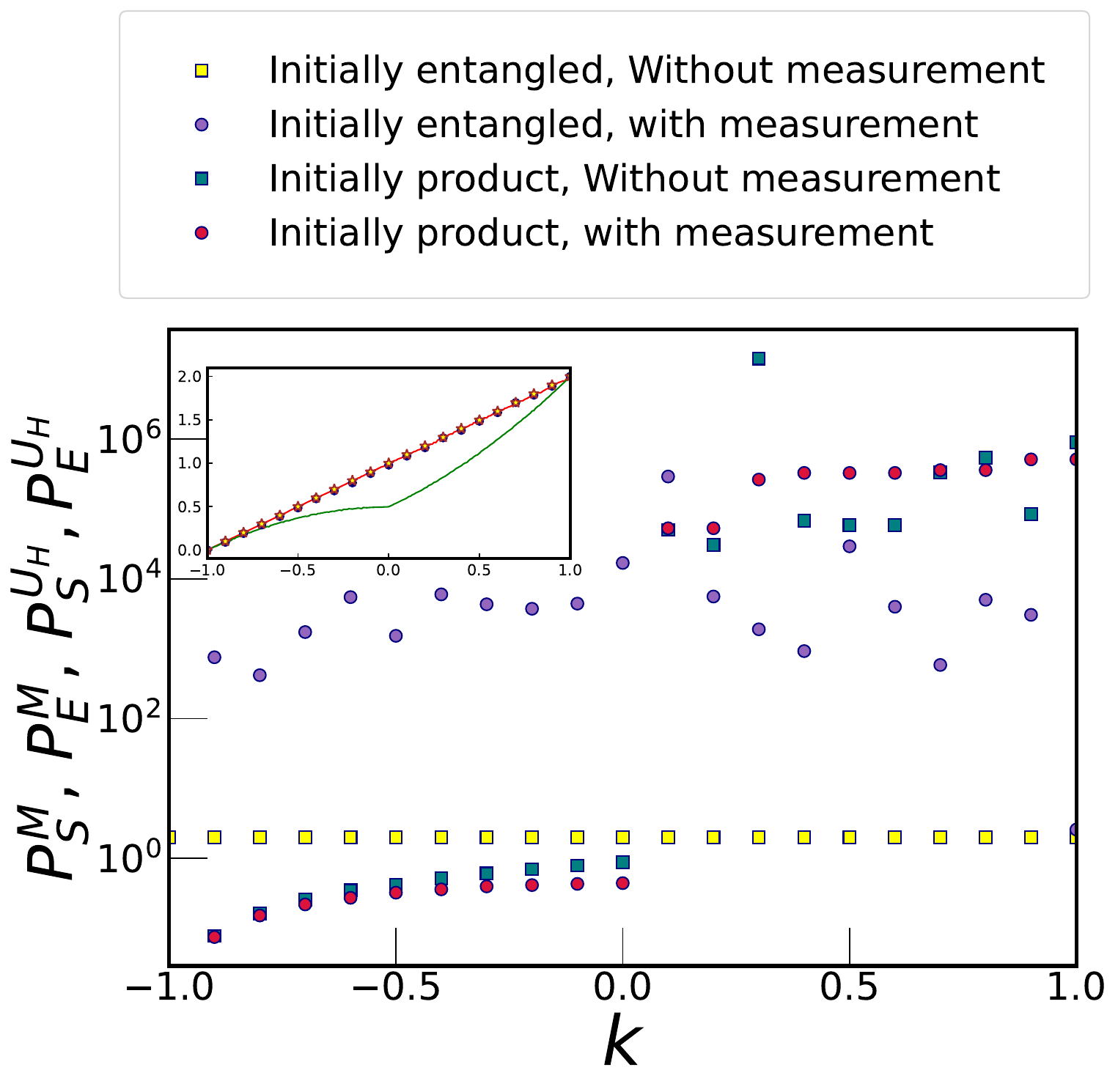}}
\caption{{In Fig.~\ref{5f}(a), we plot the maximum achievable power for both measurement-based and measurement-free processes with initial state parameter $k$. The distillable power is shown using brown and gray dots for the initial product and initially entangled states, respectively. For comparison, the power in the absence of measurement is depicted using teal and yellow boxes, corresponding to initially product states and entangled states.
In Fig.~\ref{5f}(b), the maximum probabilistic power is plotted using crimson and purple dots for the initially product and entangled states. The power for the measurement-free scenario is also included in this figure with the same colors for the initial product and entangled states as in Fig.~\ref{5f} (a). We also have included an inset in Fig.~\ref{5f} (b). In this inset plot the maximum probabilistically extractable energy for both initially entangled and product states is denoted by green and red curves. and the maximum energy in the no-measurement scenario denoted for the initial product and entangled states is denoted by blue dots and yellow stars with a brown border.}}
        \label{5f}
\end{figure}
At time $t$, we discard the auxiliary qubit and find the energy difference between the initial and time-evolved state of the battery. The maximum, $W_E^{U_H}$, of this energy difference, maximized over single qubit orthogonal states, $\{\ket{\phi},\ket{\phi_\perp}\}$, and $t$, represents the extractable energy in this method.} 
%We make a comparison between the two scenarios, and examine the case where the amount of maximum extractable energy is higher. In both the cases, the initial state of the battery is the same and parameterised by the parameter $k$.
% \textcolor{magenta}{In Fig.~\ref{figp3},  we plot $W_E^{U_H}$ and $W_S^{U_H}$ using yellow cross and blue circles  and $W^{MD}_E$ and $W^{MD}_S$ using red and green smooth curves, which are basically the distillable energies where we consider the initial state of the battery and auxiliary to be entangled and separable, in each of the two cases respectively. We can notice from the figure that $W_E^{U_H}$ 
% %and 
% is higher than $W_S^{U_H}$ 
% %are almost the same between 
% almost in the entire range $k \in [-1.0,-1.0]$, except close to the boundaries. 
% Interestingly, in contrast to the measurement-based method, here the initial entanglement between the $B$ and $A$ shows an advantage  
% %detrimental effect 
% in the energy extraction, making $W_E^{U_H}$ 
% %less 
% higher than $W_S^{U_H}$, for any fixed 
% %$k>0.5$
% $k$, which can be realized from the two dashed curves of Fig.~\ref{figp3}. However,
% %on the other hand, the probabilistically extractable 
% the distillable energy is shown to be 
% equal 
% %higher 
% in the two cases when initial entanglement is present or absent between the battery and the auxiliary for all values of $k$. }
{We calculate the maximum extractable energies, $W_E^{U_H}$ and $W_S^{U_H}$, where no measurement is involved is shown in the inset of Fig.~\ref{5f} (b) by blue dots and yellow stars. We find that the maximum extractable energy by using the global interaction, $U_H(t)$, becomes exactly equal for the initially product and entangled states for all $k$ values. This extractable energy is equal to the energy difference between the initial battery state and the ground state of the battery, which proves that there exists an optimal time and an optimal auxiliary system by which we can reach the ground state of the battery without performing any measurement. 
%It is also previously seen that distillable energy for both the initial product and entangled states is also the same with and probabilistically extractable energy for initially entangled states is equal to each other, and that is equal to the energy difference between the initial state and the ground state. But the probabilistically extractable energy for initially product state lies below that except for the initial ground state and excited state. Therefore the maximum extractable energy that can be achieved by a measurement-based procedure can also be achieved by only the global interaction without performing any measurement. This situation motivates us to examine the profile of power in both the measurement-based and the measurement-free scenarios
The same amount of energy can also be distilled from the battery, considering both the product and the entangled initial states which is shown in Fig.~\ref{non-linearity-graphx} (a), i.e., $W_{S/E}^{U_H}=W_{S/E}^{D}$. On the other hand, in the inset of Fig.~\ref{5f} (b), we plot the maximum probabilistically extractable energy for initially entangled and product states by red and green curve along with $W_E^{U_H}$ and $W_S^{U_H}$. As,   $W_{E}^{U_H}=W_{E}^{M}$, $W_{S}^{U_H}>W_{S}^{M}$, {it proves that in both the cases, i.e., initially product and initially entangled auxiliary-battery state, performing measurement does not provide any advantage over no-measurement scenario.  
However, the advantage of the performed measurement can clearly be witnessed if one checks the power.}
%using yellow crosses and blue circles, respectively, along with the distillable energies, $W^{D}_E$ and $W^{D}_S$, as red and green smooth curves. The subscripts, $E$ and $S$, in these quantities represent the relevant energies for the cases where the initial states of the battery and the auxiliary are entangled and separable, respectively. From the figure, we observe that $W_E^{D}$ exceeds $W_S^{D}$ over almost the entire range $k \in [-1.0,1.0]$, except in the vicinity of the boundaries. Notably, unlike the unitary-based approach, the presence of initial entanglement between $B$ and $A$ consistently enhances the energy extraction, rendering $W_E^{D}$ larger than $W_S^{D}$ for any fixed $k$, as evident from the two smooth curves in Fig.~\ref{figp3}. However, the unitary-based energies in the two cases, whether initial entanglement is present or absent, are found to be identical for all values of $k$. Moreover, the distillable energy is equal to the unitary-based energy when the initial state is entangled.
We define the distillable power ($P^{D}_{S/E}$) and the maximum probabilistic power ($P^{M}_{S/E}$) as
\begin{equation}\label{e19}
    P^{D/M}_{S/E}=\max_{\theta, \phi,t,\rho_{BA}(0)}\frac{\mathcal{W}^{D/M}_{S/E}(t)}{t},
\end{equation}
where $\mathcal{W}_{S/E}^{D}(t)$ and $\mathcal{W}_{S/E}^{M}(t)$ denote the distilled and probabilistically extracted energies for a given auxiliary system, measurement settings, and time, $t$ and the superscripts, $D$ and $M$, represents distillable power and maximum probabilistic power. The maximization involved in the expression of $P^{D/M}_{S}$ ($P^{D/M}_{E}$), is taken over all product (entangled) $\rho_{BA}$ that have the same marginal battery state, $\rho_B$. 
The parameters, $\theta$ and $\phi$ represent measurement directions, as defined in Eq. \ref{neweq1}. On the other hand, let $\mathcal{W}_{S/E}^{U_H}(t)$ be the extractable energy at a time $t$, using the global unitary, $U_H$, for a given auxiliary system. The subscript $S/E$ has the same meaning as in Eq.~\ref{e19}. The power for the no-measurement scenario for the initial product and entangled states is defined as 
\begin{equation}\label{e20}
    P^{U_H}_{S/E}=\max_{\theta, \phi,t,\rho_{BA}(0)}\frac{\mathcal{W}^{U_H}_{S/E}(t)}{t},
\end{equation}
where the maximization is performed over the same parameters as in Eq. \ref{e19}.
 In Fig.~\ref{5f} (a), we plot distillable power corresponding to the initially produced and initially entangled states with gray and brown dots. The quantities $P^{U_H}_{S}$ and $P^{U_H}_{E}$ are also plotted with yellow and teal colored boxes. We can see that in the range $k\in (-1,0]$, measurement-based distillable power is higher than no-measurement powers. Within this range of $k$, $P^D_S$, and $P^{U_H}_S$ are {equal}, and $P^{U_H}_E$ stays above $P^D_S$ and $P^{U_H}_S$ but below $P^D_E$. In Fig.~\ref{5f} (b), maximum probabilistic power for both the initial product and the entangled states is plotted by crimson and purple dots. $P^{U_H}_E$ and $P^{U_H}_S$ are denoted using the same point type as in Fig.~\ref{5f} (a). Here, in the range $k \in [-1,0]$, the probabilistic power for initially entangled states is higher than in the rest of the cases. Here also we note that $P^{U_H}_E$ stays higher than $P^{U_H}_S$ and $P^{M}_S$, but lower than $P^{M}_S$. Therefore, we can conclude that in this range of $k$ the measurement-based method provides benefits in terms of power. }\\
 
\section{Average energy extraction using measurements}%Average extractable energy}
\label{newsec2}
{Here, instead of post-selecting a particular outcome, we consider a probabilistic average of the energy difference between the initial and the final battery states obtained after applying joint unitary on $BA$ and acting a measurement on $A$, where the averaging is performed over all possible measurement outcomes.
%In this section, we particularly put our focus on the situation where if we average out the probabilistically extractable energy over all possible measurement operators and their outcome that is performed on the auxiliary system. 
Let the joint initial state of $BA$ be $\rho_{BA}$. After the unitary evolution, the evolved state would be $\tilde{\rho}_{BA}=U\rho_{BA}U^{\dag}$, where $U$ is any arbitrary unitary. 
We consider the measurement, performed after the unitary evolution, to be positive operator valued measurement, which may not necessarily be projective.
Let the set of these semi-positive operators be $\{M_i^\dag M_i\}_i$ which obey $\sum M_i^{\dag} M_i=I$. Hence, when the $i^{\text{th}}$ outcome occurs in the measurement, the state of $BA$ becomes $(I_B \otimes M_i\tilde{\rho}_{BA}I_B \otimes M_i^\dag)/P_i$, where $I_B$ is the identity operator acting on the battery's Hilbert space and $P_i=\Tr{(I_B \otimes M_i^{\dag} M_i\tilde{\rho}_{BA})}$ is the probability of getting the $i^{\text{th}}$ measurement outcome. %, where $i$ runs from $1$ to $N$. Here $\{\phi_i\}$ represents the set of parameters of the measurement operators.
 %Therefore the operators obey the relation, $\sum M_i^{\dag} M_i=I$. Since the measurement considered in our case is projective, the operators also satisfy $M^{2}_i=M_i$, $\forall i$. 
 Hence the energy extracted from $\rho_B$ through applying the measurement on $\tilde{\rho}_{AB}$ and selecting the $i$'th outcome is $\Delta E_i=\Tr[\rho_{B}H_B]-\Tr\left[\Tr_A(\mathbbm{I} \otimes M_i \tilde{\rho}_{BA} \mathbbm{I} \otimes M_i^{\dag})H_B\right]/P_i$.
 Hence the probabilistic average of this quantity, where the average is being taken over all measurement outcomes, can be written as
%Let us consider that the initial state of the joint system of the battery and auxiliary is given by $\rho_{BA}$ corresponding to the Hilbert space $\mathcal{H}_B \otimes \mathcal{H}_A$. After the unitary evolution, the finat state at time t is given by $\rho^{t}_{BA}=U(t)\rho_{BA}U^{\dag}(t)$.
%After that, we perform projective measurement on the auxiliary system. The measurement operators that act on the auxiliary system are given by $M(\phi_i)$, where $i=1,2,3,.....,N$. Here, $\phi_i$ are the parameters of the measurement operator. The measurement operators basically obey the properties $\sum M_i M^{\dag}_i=I$ and $M^{2}_i=M_i$. After the measurement being performed on the auxiliary qubit, the amount of probabilistically extractable energy is $P_i[\tr[\rho^t_{BA}(H_B \otimes \mathbbm{I})] -\frac{\tr(\mathbbm{I} \otimes M_i \rho^t_{BA} \mathbbm{I} \otimes M^{\dag}(H_B \otimes \mathbbm{I}))]}{P_i}$. Here $P_i$ is the probability of the outcome of a particular measurement set up. Now to do average over all possible outcomes and measurement operators, we first sum over all possible operators and outcomes and then sum over it. 
%That is,
\begin{eqnarray}
 W_{av}^M &=& \sum_i P_i \Delta E_i\nonumber\\&=&\sum_i P_i\Tr[\rho_{B}H_B]\nonumber\\&-&\sum_i\Tr\left[\Tr_A(\mathbbm{I} \otimes M_i \tilde{\rho}_{BA} \mathbbm{I} \otimes M_i^{\dag})H_B\right]\nonumber\\
    &=& \Tr[\rho_{B}H_B ]\nonumber \\
    &-&\sum_{i}\Tr\left[I_B \otimes M_i \tilde{\rho}_{BA} I_B \otimes M_i^{\dag}(H_B \otimes I_A)\right]\nonumber \\
&=& \Tr[\rho_{B}H_B ]-\Tr\left[\tilde{\rho}_{BA}\left(H_B \otimes \sum_iM_i^{\dag}M_i\right)\right]\nonumber \\
&=&\Tr[\rho_{B}H_B ]- \Tr[\tilde{\rho}_{BA}(H_B \otimes I_A)]\nonumber\\
&=&\Tr[\rho_{B}H_B ]-\Tr[\tilde{\rho}_{B}H_B],
\end{eqnarray}
where $\tilde{\rho}_B$ and $I_A$ are the state of the battery after the unitary evolution and the identity operator which acts on the auxiliary's Hilbert space. Therefore, the average extractable energy is independent of the performed measurement and is same as the extractable energy using the global unitary. This is what we expect from the no-signaling theorem: that the measurement performed on the auxiliary will not affect, on average, the energy of the battery. Hence the behavior of $W_{av}^M$ is the same as $W^{U_H}$ when we consider $U=U_H(t)$.}

{\section{Extractable energy and power from larger batteries}
\label{newsec3}}
{In this section, we firstly want to examine if the single qubit auxiliary can still maintain its impact if we increase the size of the battery, in the context of {distillable, maximum probabilistically extractable} energy, and also compare it with the extractable energy of the no measurement scenario. In this regard, we consider an $N$ qubit battery, the initial state of which is $\rho_{B_N}=\rho^{\otimes N}_{B}$,
%In this section we wish to study the variation of probabilistically extractable energy with the change of dimension of the system and keeping the dimension of the auxiliary system fixed. In that regard, we take the initial state of the battery as $\rho_B=\rho^{\otimes n}_{Bn}. $ 
where $\rho_{B}=\begin{bmatrix}
\frac{1}{2} & 0 \\
0 &\frac{1}{2}\\
\end{bmatrix}$. 
Here both $\rho_{B_N}$ and $\rho_{B}$ are maximally mixed states, implying that they are passive. We keep the auxiliary system as single qubit, the state of which is presented in Eq.~\eqref{auxi} of Sec.~\ref{mp}. The Hamiltonians describing the energy of the battery and the auxiliary are, respectively, 
$$H_B^N=h\sum_{i=1}^{N}\sigma_i^z\text{ and }H_A=h\sigma_{N+1}^z.\label{4eq1}$$ 
 we switch on an interaction between the $N+1$ qubits which is described by the following interaction Hamiltonian:
\begin{equation}
       H_{I}^{N+1}=J\sum_{i=1}^{N+1} \sigma^i_x \sigma^{i+1}_x.
  \end{equation}
    Hence the total Hamiltonian governing the unitary evolution is given by
  \begin{equation}
       H_{tot}=H_B^N+H_A+H_I^{N+1}=J\sum_{i=1}^{N+1} \sigma^i_x \sigma^{i+1}_x+h\sum_{i=1}^{N+1}\sigma^i_z.
  \end{equation}
}

{We follow the same measurement-based energy extraction process, i.e., we act as a joint unitary, $U_H^{N+1}$, on the $N$ qubit battery-auxiliary system, then perform a projective measurement on the auxiliary, and calculate the distillable energy and maximum probabilistically extractable energy by maximizing over time measurement basis and auxiliary system. In the measurementless scenario, the protocol is exactly the same, but no measurement is performed on the auxiliary qubit, and it is optimized over the auxiliary system and time. { Here for the numerical study we chose $J=2$ and $h=1$.
We obtain the distillable energy and maximum probabilistically extractable energy by maximizing over the measurement basis, auxiliary system, and time denoted by $W^D_N$, $W_N^M$, and $W^{U_H}_N$. In the inset of Fig.~\ref{fig6}, we plot $W^D_N$, $W_N^M$, and $W^{U_H}_N$ with $N$, where $W^D_N$, $W_N^M$, and $W^{U_H}_N$ are denoted by orange and blue dots with navy blue borders and yellow stars with a navy blue border. Here we can see that $W^D_N$, $W_N^M$, and $W^{U_H}_N$ decrease with the increase of $N$. We can see that no-measurement scenario provides a better advantage over the measurement-based case. This particular scenario leads us to check the status of distillable power, maximum probabilistic power and maximum power in no measurement scenario, which is denoted by $P^D_N$, $P_N^M$, and $P^{U_H}_N$ and plotted with light color, wheat-colored hexagons, and dark slate blue-colored stars with navy blue borders in the main figure of Fig.~\ref{fig6}. Here, one can note that although $P_N^M$ lies below $P_N^{U_H}$, the distillable power $P_N^D$ always remains higher than $P_N^{U_H}$ for all the $N$ we have considered. It depicts a clear advantage of measurement-based protocol over the non-measurement case.} \\
%

%basically maximally mixed state, and $\rho_B$ is the maximally mixed state in higher multiqubit dimension. The auxiliary system remains as the single qubit state, which is referred to in Sec. \ref{mp} in Eq. \ref{auxi}. On the other hand, the total Hamiltonian of the system and auxiliary is modified by
 	\begin{figure}
		\centering
		\includegraphics[scale=0.28]{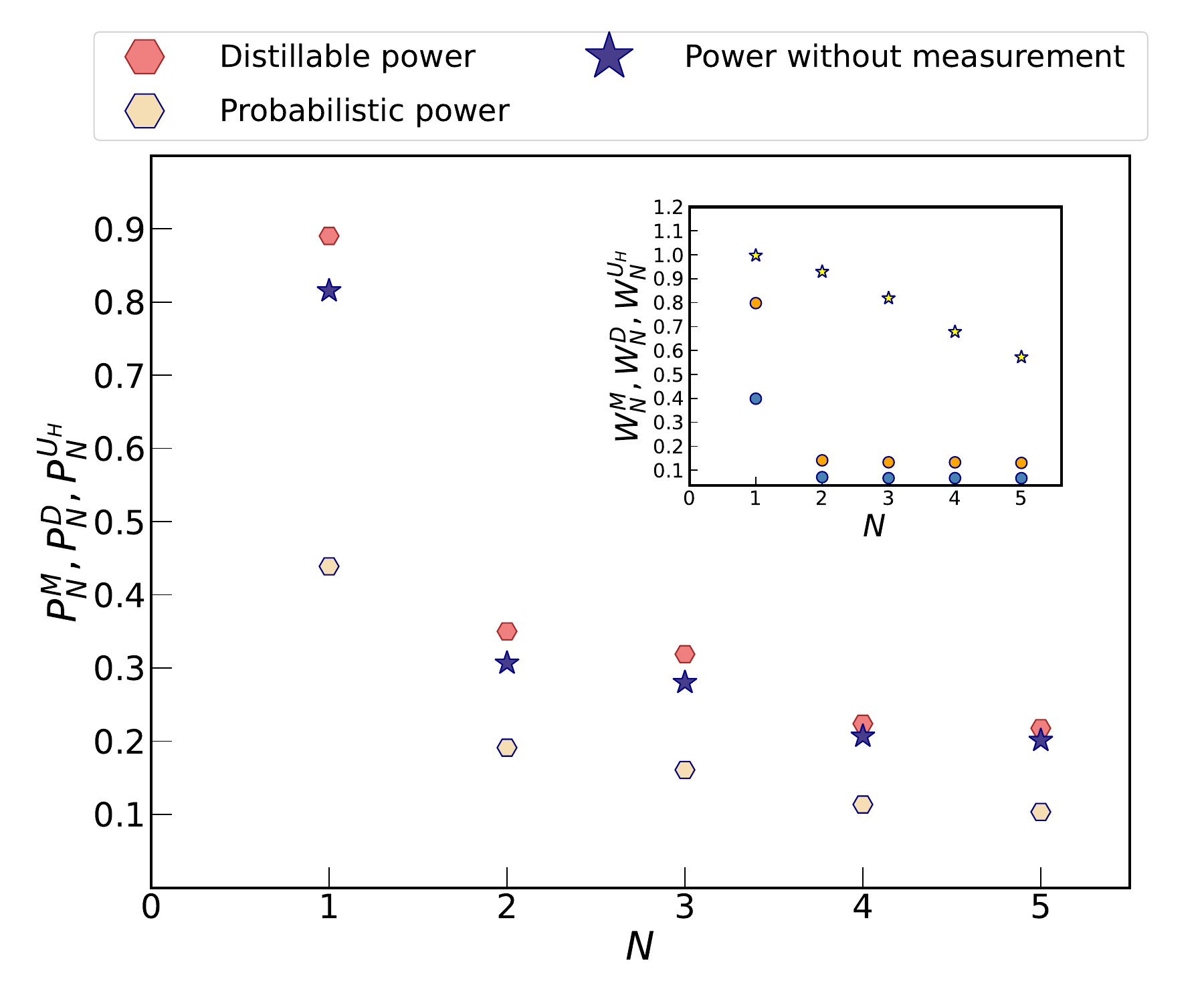}
		\caption{{The distillable power ($P_N^D$), probabilistic power ($P_N^M$), and power without measurement scenario ($P_N^{U_H}$) of the battery are plotted along the vertical axis in light coral and wheat-colored hexagons and dark slate blue-colored stars with navy blue stars, respectively, versus the number of qubits ($N$) of the battery system along the horizontal axis, where the auxiliary system is kept to be a single qubit. At the same time we plot distillable, maximum probabilistic energy and the maximum energy where no measurement is involved with the number of qubits in the battery $N$. Each of the quantities that is denoted by $W^D_N$, $W^M_N$, and $W^{U_H}_N$ and marked with orange and blue dots with navy blue borders and yellow stars with navy blue borders.} }
  \label{fig6}
	\end{figure}

  }

\section{measurement-passive states}
\label{3}
Analogous to the notion of passive states in unitary extraction of energy, we can define measurement-passive states as the states from which no energy can be extracted using the measurement-based technique discussed in the preceding section.

In this section, considering the scenario where the system and auxiliary qubits are initially prepared in a separable state, we want to determine the MPS corresponding to the Hamiltonian, {$H_{BA}$}, expressed in Eq. \eqref{4eq5}. 
%The probabilistic energy extraction from the state $\rho_S$ can be calculated following the same measurement-based protocol. 
{In particular, we want to find the states for which $W^D_S=0$. Since the ground state is the lowest energy state of the corresponding Hamiltonian, it follows that the ground state is trivially an MPS. We will prove that, except for the ground state, there does not exist any state for which $W^D_S=0$. In this regard, we will first show that for all states (except the ground state) $W^P_S\neq 0$. Since $W^D_S\geq W^M_S\geq 0$, $W^P_S\neq 0$ would directly imply $W^D_S\geq 0$.} The expression of $W^M_S$ consists of a maximization over the initial auxiliary state, i.e., over $(r,\theta,\phi)$, the total time of unitary evolution, $t$, and the measurement-basis, parameterized by $(\theta_1,\phi_1)$. %Instead of delving into the complexities of that optimisation, we will prove that every state except the ground state can provide non-zero amount of energy. 
We will now prove that corresponding to every state except the ground state, there exists at least one set of parameters, $\{r,\theta,\phi,t,\theta_1,\phi_1\}$, such that the measurement-based protocol can provide a non-zero amount of energy. In other words, the measurement-based procedure does not have any non-trivial MPS.

Let the two-qubit battery-auxiliary system be initially prepared in a product state of the form $\rho^s_{BA}=\rho_{B} \otimes \rho_{A}$. We consider the state of the battery to be a general single-qubit state,  $\rho_B=\frac{1}{2}(I_2+\vec{s}.\vec{\sigma})$, which can be represented by a point in the Bloch sphere having Cartesian coordinate  $(s\cos{\tilde{\theta}}\sin{\tilde{\phi}},s\sin{\tilde{\theta}}\sin{\tilde{\phi}},s\cos{\tilde{\theta}})$. Here $\tilde{\theta}$ and $\tilde{\phi}$ are respectively the zenith and azimuthal angles of the point, and $s$ is its distance from the center of the Bloch sphere.  The point parameterized by $s=1$, $\tilde{\theta}=\pi$, $\tilde{\phi}=0$ represents the ground state of the Hamiltonian, $H_B=h\sigma_z$ for $h>0$. %$\rho_A$ has the same form as given in Eq. \eqref{8}. 
%Consider the case where a global unitary $U$ is allowed to govern the evolution of the combined state of the battery and auxiliary given by in \eqref{3}. Under unitary evolution, the state transforms to $U\rho_{SA}U^{\dagger}$. Finding the set of states from which no energy can be taken is the goal here. This is accomplished by demonstrating the set of states from which energy can be extracted for any non-zero interaction strength,  and the complement of that set will imply the set of passive states. 
$\rho_A$ is considered to be $\kett{1}\braa{1}$.
We turn on the interaction $\bar{H}_{BA}$ (of Eq.~\eqref{4eq5}) between the battery and the auxiliary for a fixed amount of time, say $t$. After time $t$, we perform a projective measurement on the auxiliary qubit in the $\sigma_z$ eigenbasis, $\{\kett{0},\kett{1}\}$. We select the outcome corresponding to the measurement operator $\kett{1}\braa{1}$. The probabilistically extracted energy in this procedure is given by
 \begin{eqnarray}
   W_P =\frac{h}{4 \left(4 h^2 + J^2\right)}\Big{[}-4 h^2 + \left(4 h^2 + J^2\right) \cos(2 J t)\nonumber \\- 
   J^2 \cos\left(2 \sqrt{4 h^2 + J^2} t\right)\Big{]} \left(-1 + 
   s^2 \cos^2\tilde{\theta}\right).
   \label{9}
 \end{eqnarray}
Showing that for $h>0$, except for the ground state and first excited state, i.e., for $s=1$ and $\tilde{\theta}=\pi$ and $s=1$ and $\tilde{\theta}=0$, $W_{P}$ is positive for all $s$ and $\tilde{\theta}$. To see this, we essentially examine the behavior of $W_P$ for $t$ close to $0$. The lowest order non-zero term in the series expansion of $W_P$ about $t=0$ is given by
 \begin{equation}
   W_{P} =\frac{-8 h(4 h^4 J^2 + h^2 J^4) (-1 + 
   s^2 \cos^2\tilde{\theta}))}{12(4 h^2 + J^2)}t^4.
   \label{10}
 \end{equation}
The factor, $\left(-1 +s^2 \cos^2\tilde{\theta}\right)$, is always negative except for the ground and excited states of the battery's Hamiltonian, $H_B$, in which case the factor becomes equal to zero. All the other factors in the expression of $W_{P}$, i.e., $h$, $(4 h^4 J^2 + h^2 J^4)$, and $(4 h^2 + J^2)$, are surely positive for all positive values of $h$. On the other hand, if $h<0$, we can choose the initial auxiliary system as $\ket{0}\bra{0}$ and choose the outcome corresponding to the measurement operator $\ket{0}\bra{0}$. The lowest order term for maximum probabilistically extractable energy expended around $t=0$ would be 
    \begin{equation}
   W_{P} =\frac{8 h(4 h^4 J^2 + h^2 J^4) (-1 + 
   s^2 \cos^2\tilde{\theta}))}{12(4 h^2 + J^2)}t^4.
   \label{10}
 \end{equation}
 Similarly, as $h$ and $\left(-1 +s^2 \cos^2\tilde{\theta}\right) $ is negative except for ground state and excited state and $(4 h^4 J^2 + h^2 J^4)$ and $(4 h^2 + J^2) $ are positive but $h$ is also negative; it indicates that energy can be extracted probabilistically from all the states except the ground and excited states.

 As a result, we see that for any non-zero value of $h$ and for any state $\rho_B$, except for the eigenstates of $H_B$, if the measurement is done after a very small evolution time $t$, the probabilistically extracted energy is positive. This almost completes our proof.
   The ground state of $H_B$ is trivially passive. So we only need to demonstrate that there exists a time, $t$, an auxiliary state, $\rho_A$, and a measurement-basis, $\{\kett{\phi},\kett{\phi_{\perp}}\}$, such that $W_{P}>0$ for the excited battery state, $\rho_B=\kett{0}\braa{0}$. %We demonstrate analytically that energy can be extracted even if $J$ is very small but not equal to zero for all the states aside from ground state and excited state because for ground state and excited state the probabilistically extracted energy according to \eqref{10} is zero. This is because we are focusing on this particular auxiliary system and the specific outcome of the computational measurement operator set.
In this regard, let us consider the initial auxiliary state to be $\rho_A=\kett{0}\braa{0}$. Again we perform measurement on the auxiliary qubit after a fixed time $t$ on the same measurement-basis $\{\kett{0},\kett{1}\}$. If we consider the battery state corresponding to the output $\kett{1}\braa{1}$, the amount of probabilistically extracted energy in this process would be   \begin{equation}
   W_P ={2hJ^2}\frac{ \sin[(\sqrt{4 h^2 + J^2)}t]}{4 h^2 + J^2}.
   \label{12}
 \end{equation}
Corresponding to every $h$ and $J$, we can find a suitable $t$ for which the above quantity is positive. So a non-zero amount of energy can be successfully extracted from the state $\rho_B=\kett{0}\braa{0}$ as well. {As we mentioned before, $W_P>0$, for a particular set of parameters, implies $W_S^M>0$ since the latter involves a maximization over all relevant parameters. Moreover, since by definition $W_S^D$ includes sum over positive terms and $W_S^M$ invloves a single positive term, $W_S^M>0$ implies $W_S^D>0$. Hence we realize the Hamiltonian's ground state is the sole measurement-passive state that we have.}

{We compared three methods of energy extraction. The first one is the maximum extractable energy using unitary operations, also referred to as the ergotropy, while the second and third methods of energy extraction involve a measurement applied on an auxiliary qubit attached to the battery, after allowing them to interact for a certain amount of time, considering the initial battery-auxiliary state, before the mentioned interaction to be uncorrelated and correlated, respectively. Though in all three cases, the fully charged state is the excited state of the relevant Hamiltonian, we found the empty state, which refers to the state from which no energy can be extracted, differs from one protocol to the other. In the measurement-based protocols, the ground state of the relevant Hamiltonian serves to be the only empty or chargeless state.  However, for the unitary extraction process, the empty state depends on the condition mentioned in Sec.~\ref{22}, which shows that the ground state is not the only passive state.

%One can notice from both Eqs. \eqref{10} and \eqref{12} that the condition $J=0$ implies $W_\mathcal{P}=0$. Actually, not only does $W_\mathcal{P}$ become zero in this case, but even if we maximize $W_\mathcal{P}$ over all possible initial states of the auxiliary qubits, time $t$, and measurement basis, the maximum extractable energy, $W^M_S$, will still be zero. This happens because $J=0$ refers to no interaction between battery and auxiliary, and if the initial state jointly describing $B$ and $A$ is a product, so will be the final state. Therefore, doing any measurement on the auxiliary qubit and post-selecting any appropriate state will not affect the battery. Moreover, since in the absence of the interaction, the evolution within the time $t$ reduces to a local evolution with respect to their Hamiltonians, which describe the self-energies of the battery and auxiliary, that evolution will also not change the energy of the battery.

\section{Conclusion}
\label{concl}
A conventional approach to harnessing energy from quantum batteries involves the utilization of unitary operations. The ergotropy of a quantum battery refers to the maximum amount of energy that can be extracted from it through unitary dynamics. Once all the accessible energy has been extracted, the battery reaches a passive state, beyond which no further energy can be obtained via unitary operations. In the case where the initial state of the battery is pure, the passive state with respect to the unitary operation that it attains after providing all of its energy is the ground state of the associated Hamiltonian. However, it is important to note that for a given Hamiltonian, there exist infinitely many (mixed) passive states from which energy extraction using unitary operations is not possible.

{In this paper, we considered energy extraction using measurements. In particular, we introduced an auxiliary system, on which we make the measurement, for energy extraction. Initially, the auxiliary qubit interacts with the battery for a certain amount of time, after which a measurement is performed on the auxiliary qubit. {A single or multiple preferred outcomes of the measurement are selected, which results in successful energy extraction}. {We defined the product of the difference between the initial and final energies of the battery, corresponding to a single selected outcome, and the probability of getting that outcome as probabilistically extractable energy. The sum of the product of the probability associated with the preferred measurement outcome that provide positive energy,
and the corresponding energy difference between the initial
and final states of the battery corresponding to that outcome, was referred to as the distilled energy. The distillable energy is the optimal value of this quantity,  maximized over the initial auxiliary state, time of performing measurement, and measurement basis.}  Our main motive is to draw a comparison between distillable or maximum probabilistically extractable energy with the battery's ergotropy. In this regard, we investigated two scenarios: in the first case, the initial state of the battery and the auxiliary is taken to be a product, and in the second case, it is considered to be entangled. Exploring the situations, we showed that the measurement-based protocol provides significantly better energy extraction from a quantum battery than the same based on unitary operations. By comparing the two mentioned scenarios, we realize that the distillable energy is the same in both the cases, whereas the presence of entanglement provides an enhancement in the maximum probabilistically extractable energy, maximized over time of interaction, measurement operators, and auxiliary state. We scrutinized the optimal parameters for which the distillable energy can be found through measurement. We showed that if, instead of selecting particular outcomes, the average over all outcomes is considered, then the resulting quantity, i.e. the average extractable energy, becomes independent of the measured.  
{Moreover, we found that the measurement-based method does not surpass the global unitary approach in terms of extractable energy. It yields the same energy in certain cases (e.g., energy distillation with either entangled auxiliary-battery states, and probabilistic energy extraction with entangled auxiliary-battery states), and even less energy in others (specifically, when the auxiliary-battery state is a product state under probabilistic extraction) cases compared to the no-measurement case. However, the measurement-based method generally provides a significant advantage in terms of power, particularly when the initial battery-auxiliary states are to be entangled.}
Furthermore, we found, this enhancement in power through performance of  (distillable power), though decreases with increasing size of the battery, still remains higher than the measurement less scenario. Finally, we determined that the measurement-based energy protocols can extract energy from any battery state, except the ground state, which indicates their superiority over unitaries acting only on the battery.}

}

%Finally, we showed that if a non-zero amount of interaction can be switched on between the auxiliary and the battery for an infinitesimal time-interval, energy can be extracted from any battery state except the ground state. Therefore, the set of measurement-passive states has only one element, viz., the ground state.
%Therefore, in the case where the system and auxiliary are initially separable, and an interaction between them is switched on after some time, there is no other passive state except the ground state of the relevant Hamiltonian.
\acknowledgments
    We acknowledge computations performed using Armadillo. The research of KS was supported in part by the INFOSYS scholarship. We also acknowledge partial support from the Department of Science and Technology, Government of India through the QuEST grant (grant number DST/ICPS/QUST/Theme- 3/2019/120). 
  % Switch to single-column mode

\appendix
\setcounter{figure}{0}

\section*{Appendix}

\section{Initial settings providing optimal energy extraction} \label{appA}

In Table~\ref{tab:updated_params}, we present one set of optimal parameters representing measurement basis, time of evolution, and initial auxiliary state, for different initial configurations of $B$ defined through $k$ for which the distillable energy can be extracted through measurements.
However, it should be noted that there are a large number of measurement settings, auxiliary states, and evolution times, using which we can squeeze out the distillable energy from a given initial state of the battery. 
 For instance, for a given initial 
 state, a
 $\rho_B=\begin{bmatrix}
			0.7 & 0 \\
			0 & 0.3 \\
		\end{bmatrix}$, and initially product battery-auxiliary system both choices of parameter settings, that are, $\{\theta=2.22641, \phi=5.24603, r=0.944272, t=0.725847, \theta_1=1.36077, \theta_1=5.68614 \}$ and $\{\theta=1.15846, \phi=3.33632, r=0.391733, t=0.104449, \theta_1=1.74971, \phi_1=0.18532\}$ can extract the entire distillable energy from $\rho_B$. Here, $\{\theta, \phi\}$, $t$, and $\{\theta_1, \phi_1, r\}$ are the parameters defining the measurement operator, time of evolution, and initial auxiliary state, respectively.  Further, we find that one can distill energy from the battery using the measurement-based method even when the time of evolution is as small as 0.01$\frac{\hbar}{h}$.
         \begin{figure*}[]
    \subfigure[]{\includegraphics[width=0.40\textwidth]{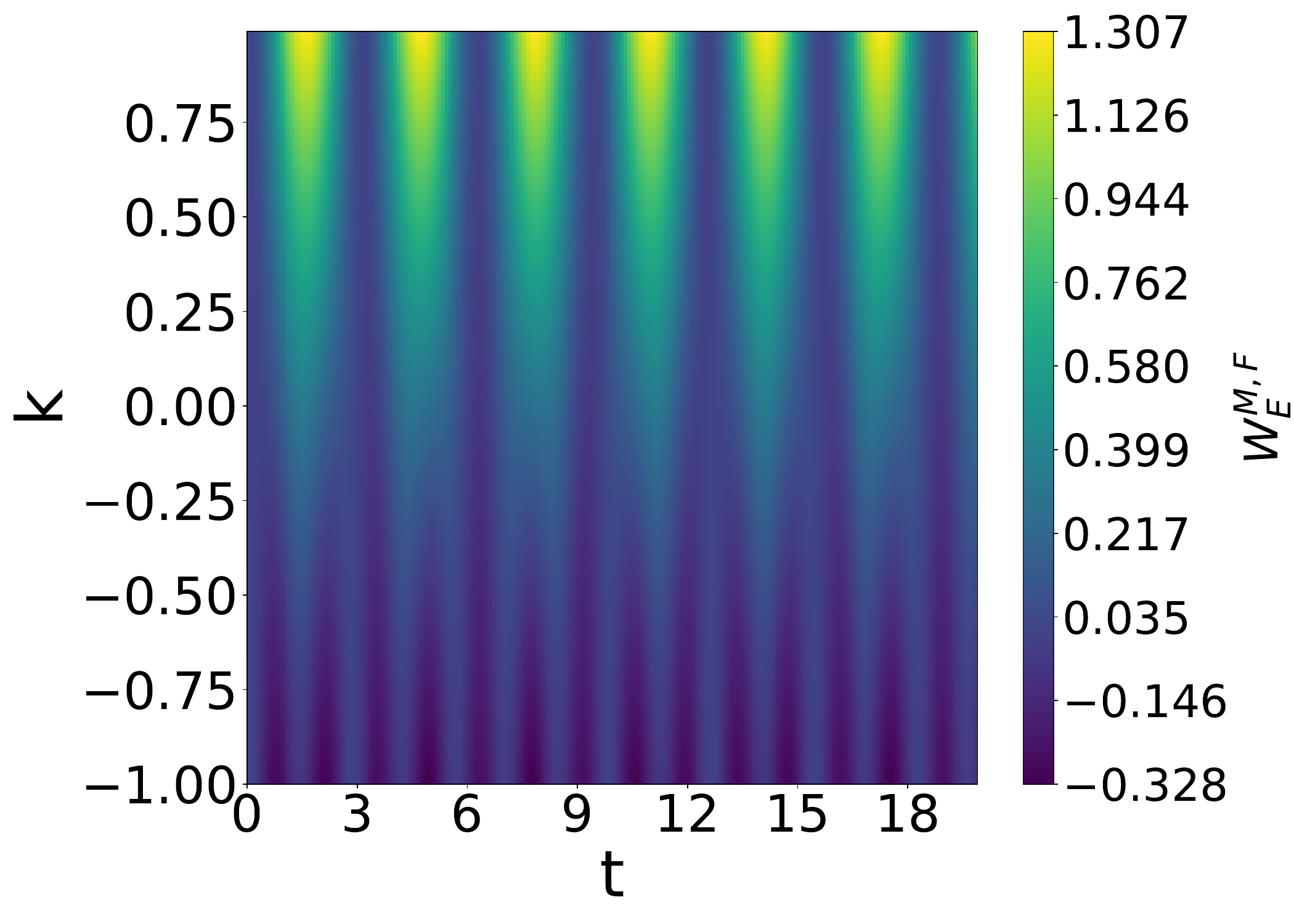}}
    \hspace{0.32cm}
        \subfigure[]{\includegraphics[width=0.40\textwidth]{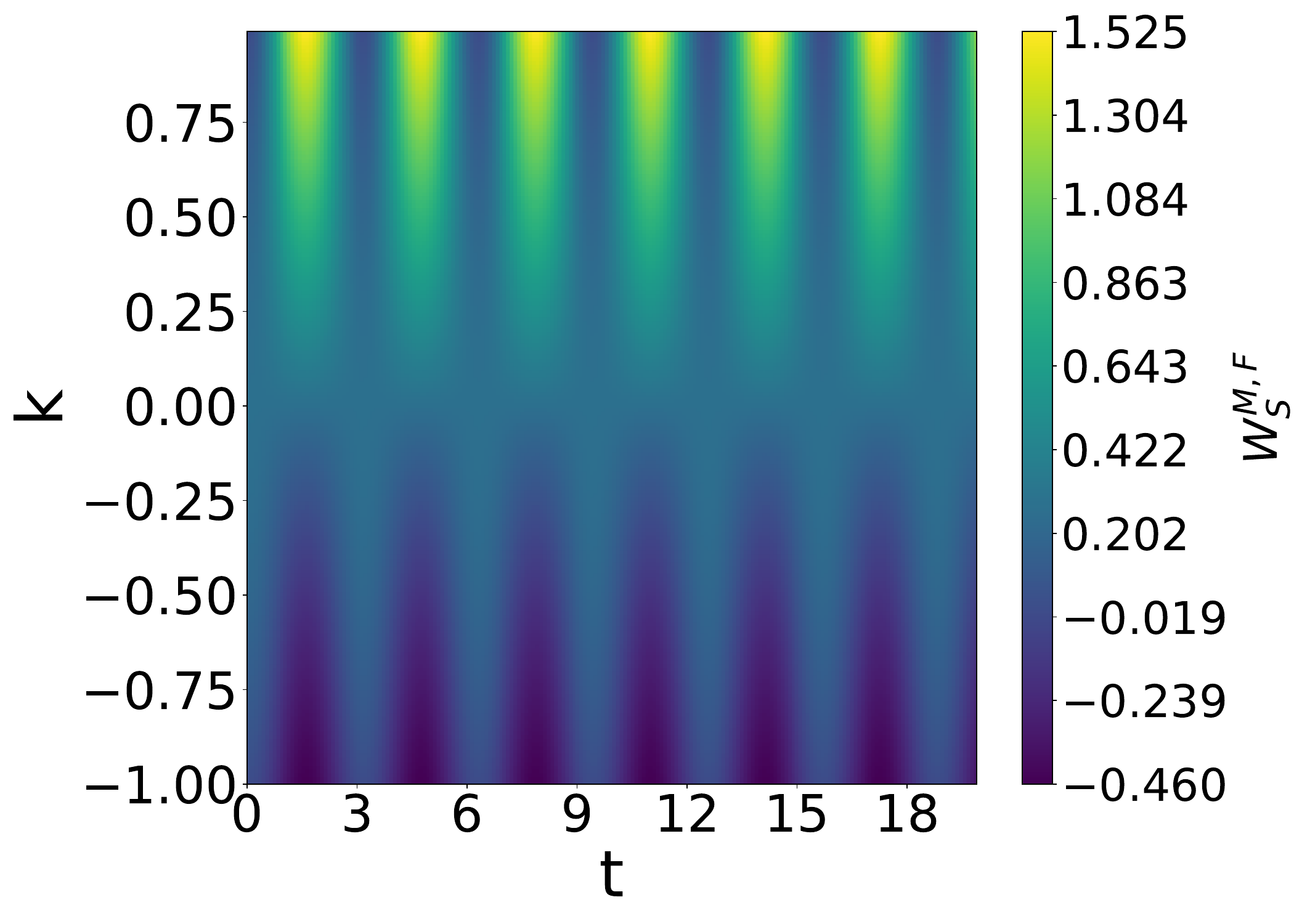}}
\caption{{Fig.~\ref{1a} (a) shows the probabilistically extractable energy $W^{M,F}_S$ in a two-dimensional projection for a fixed measurement setting and auxiliary system (as described in subsection~\ref{2xb}). Here, the initial state parameter $k$ is plotted along the vertical axis and the interaction time $t$ through the Hamiltonian $H_{BA}$ along the horizontal axis, assuming the battery and auxiliary start in a product state. Fig.~\ref{1a} (b) presents the corresponding results for initially entangled states, with the horizontal and vertical axes defined as in Fig.~\ref{1a} (a). In this case, the probablistically extractable energy is denoted as $W^{M,F}_E$. The color bars in the respective plots indicate the values of $W^{M,F}_S$ and $W^{M,F}_E$ for given $t$ and $k$.}}
        \label{1a}
\end{figure*}
        %Here we have given a example of a choice of optimal measurement, optimal auxiliary and interaction time.
%                  \begin{figure}[]
%     \subfigure[]{\includegraphics[width=0.47\textwidth]{3D_sepF.pdf}}
    
%         \subfigure[]{\includegraphics[width=0.47\textwidth]{3D_entF.pdf}}
% \caption{\textcolor{blue}{Fig.~\ref{1a} (a) shows the distillable energy $W^{M,F}_S$ in a two-dimensional projection for a fixed measurement setting and auxiliary system (as described in subsection~\ref{2xb}). Here, the initial state parameter $k$ is plotted along the $Y$ axis and the interaction time $t$ through the Hamiltonian $H_{BA}$ along the $X$ axis, assuming the battery and auxiliary start in a product state. Figure~\ref{1a} (b) presents the corresponding results for initially entangled states, with the $X$ and $Y$ axes defined as in Fig.~\ref{1a} (a). In this case, the distillable energy is denoted as $W^{M,F}_E$. The color bars in the respective plots indicate the values of $W^{M,F}_S$ and $W^{M,F}_E$ for given $t$ and $k$.}}
%         \label{1a}
% \end{figure}
\begin{table}
    \centering
    \begin{tabular}{|c|c|c|c|c|c|c|}
        \hline
        $k$ & $\theta$ & $\phi$ & $r$ & $t$ & $\theta_1$ & $\phi_1$ \\
        \hline
        -1.0  & 2.145 & 0.440 & 0.372 & 17.316 & 0.706 & 0.751 \\
        -0.9  & 1.476 & 1.230 & 0.158 & 12.131 & 2.838 & 4.595 \\
        -0.8  & 1.868 & 0.156 & 0.066 & 27.756 & 0.124 & 5.160 \\
        -0.7  & 1.200 & 0.946 & 0.852 & 11.922 & 2.257 & 2.721 \\
        -0.6  & 2.612 & 3.600 & 0.516 & 20.284 & 2.953 & 3.560 \\
        -0.5  & 1.944 & 4.390 & 0.301 & 25.747 & 1.943 & 1.121 \\
        -0.4  & 2.336 & 3.316 & 0.209 & 6.937 & 2.371 & 1.686 \\
        -0.3  & 0.646 & 0.379 & 0.238 & 14.767 & 1.093 & 5.256 \\
        -0.2  & 2.805 & 1.910 & 0.100 & 26.783 & 0.164 & 6.103 \\
        -0.1  & 0.055 & 0.836 & 0.008 & 4.330 & 0.591 & 0.385 \\
         0.0  & 1.468 & 3.490 & 0.671 & 34.388 & 1.287 & 1.225 \\
         0.1  & 1.860 & 2.416 & 0.579 & 34.199 & 1.715 & 1.790 \\
         0.2  & 0.170 & 5.762 & 0.609 & 18.564 & 0.437 & 5.360 \\
         0.3  & 0.562 & 4.688 & 0.516 & 18.763 & 0.865 & 5.925 \\
         0.4  & 2.014 & 1.751 & 0.546 & 14.725 & 2.729 & 3.212 \\
         0.5  & 2.406 & 0.678 & 0.453 & 26.768 & 0.015 & 3.777 \\
         0.6  & 2.758 & 5.195 & 0.995 & 29.943 & 2.416 & 1.611 \\
         0.7  & 0.009 & 4.121 & 0.903 & 22.660 & 2.843 & 2.176 \\
         0.8  & 0.400 & 3.047 & 0.811 & 25.033 & 0.129 & 2.742 \\
         0.9  & 0.202 & 4.149 & 0.235 & 12.932 & 1.884 & 0.046 \\
         1.0  & 0.594 & 3.075 & 0.143 & 16.496 & 2.311 & 0.611 \\
        \hline
    \end{tabular}
    \caption{The table depicts the optimal parameter set of measurement operator $\{ \theta, \phi \}$, interaction time $t$, and auxiliary system $\{ \theta_1, \phi_1, r \}$ corresponding to different input states indicated by distinct values of $k$.}
    \label{tab:updated_params}
\end{table}

%\label{newsec1}

\label{4}

\bibliography{main}
\end{document}